\RequirePackage{fix-cm}
\documentclass[smallextended,natbib]{svjour3}
\smartqed  
\usepackage{graphicx}
\usepackage{url}
\usepackage{hyperref}
\usepackage{color}
\usepackage{siunitx}

\usepackage{graphbox}
\usepackage[font=footnotesize,labelfont=bf]{caption}

\journalname{}

\newcommand{\st}[1]{\mathrm{start}(#1)}
\newcommand{\en}[1]{\mathrm{end}(#1)}

\newcommand{\PI}{{\rm\Pi}}

\newcommand{\PCY}{{\mbox{PCY}}}

\newcommand{\N}{\mathcal{N}}

\begin{document}

\title{Is science driven by principal investigators?\thanks{\textcolor{red}{This a final preprint version; the final version of the manuscript can be found at \url{https://doi.org/10.1007/s11192-018-2900-x}}\newline This research was partially supported by Slovenian Research Agency Program P1--0383 and Projects J1--8155, N1--0057.}
}


\author{Andrej Kastrin \and Jelena Klisara \and Borut Lu\v{z}ar \and Janez Povh}


\institute{A. Kastrin \at
University of Ljubljana, Faculty of Medicine, Institute for Biostatistics and Medical Informatics, Slovenia.\\
\email{andrej.kastrin@mf.uni-lj.si}           
\and
J. Klisara \at
University of Ljubljana, Faculty of Computer and Information Science, Slovenia.\\
\email{jelena.klisara@fri.uni-lj.si}           
\and
B. Lu\v{z}ar \at
Faculty of Information Studies in Novo mesto, Slovenia \& Pavol J. \v{S}af\'{a}rik University, Faculty of Science, Ko\v{s}ice, Slovakia.\\
\email{borut.luzar@gmail.com}           
\and
J. Povh \at
University of Ljubljana, Faculty of Mechanical Engineering, Slovenia.\\
\email{janez.povh@fs.uni-lj.si}           
}

\date{Received: date / Accepted: date}

\maketitle

\begin{abstract}\small 
In this paper we consider the question what is the scientific and career performance of principal investigators (PI's) of publicly funded research projects compared to scientific performance of all researchers. Our study is based on high quality data about (i) research projects awarded in Slovenia in the period 1994--2016 (\num{7508} projects with \num{2725} PI's in total) and (ii) about scientific productivity of all researchers in Slovenia that were active in the period 1970--2016---there are \num{19598} such researchers in total, including the PI's. We compare average productivity, collaboration, internationality and interdisciplinarity of PI's and of all active researchers. Our analysis shows that for all four indicators the average performance of PI's is much higher compared to average performance of all active researchers. Additionally, we analyze careers of both groups of researchers. The results show that the PI's have on average longer and more fruitful career compared to all active researchers, with regards to all career indicators. The PI's that have received a postdoc grant have at the beginning outstanding scientific performance, but later deviate towards average. On long run, the PI's leading the research programs (the most prestigious grants) on average demonstrate the best scientific performance. In the last part of the paper we study \num{23} co-authorship networks, spanned by all active researchers in the periods 1970--1994, \ldots, 1970--2016. We find out that they are well connected and that PI's are well distributed across these networks forming their backbones. Even more, PI's generate new PI's, since more than $90\%$ of new PI's are connected (have at least one joint scientific publication) with existing PI's. We believe that our study sheds new light to the relations between the public funding of the science and the scientific output and can be considered as an affirmative answer to the question posed in the title.

\end{abstract}

\section{Introduction}
\label{sec:introduction}

\subsection{Motivation}
Making science metrics more rigorous is getting more and more important not only for scientists themselves, who want to be evaluated by correct measures, but also for other stakeholders in the research ecosystem~\citep{Lane2010}. This is of particular importance for the public entities which are in charge of providing public resources for science. They want to be sure that the money spent for science pays back.

For example, European Commission shares in the recent Horizon 2020 work programme 2018--2020---Strategic Programme Overarching Document\footnote{\url{http://ec.europa.eu/programmes/horizon2020/sites/horizon2020/files/stratprog_overarching_version_for_publication.pdf}} the following belief: ``EU funded research and innovation draws on the world beating excellence of EU universities, 
research performing and innovative companies including small and mid-sized firms, and the centres of expertise \ldots to give an increased bang for every Euro spent''.

Slovenia, as an EU member state, has positioned excellent science in the centre of ``Resolution on Research and Innovation Strategy of Slovenia 2011--2020''\footnote{\url{http://www.arhiv.mvzt.gov.si/fileadmin/mvzt.gov.si/pageuploads/pdf/odnosi_z_javnostmi/01.06.2011_dalje/01.06._RISSdz_ENG.pdf}}. This document proposes several measures to achieve excellent and internationally visible research, including increasing public funding of research, together with key performance indicators for convergence to the strategic goals. Most of them are related to production of high quality publications, with an additional focus to scientific publications co-authored with foreign researchers.

We mention this examples because both demonstrate belief that more money and more enhanced measures will eventually imply better and more productive science in terms of better publications, more international collaboration, more research projects, and more citations. Research, especially basic, is usually funded by public money which is always limited, therefore for better governance of scientific stakeholders and for more efficient spending of public money, a deeper understanding of relations between funding and scientific performance is needed. Our paper provides partial answer to this need, at least for the case of Slovenia.

Publicly funded scientific projects are a standard lever to steer scientific community towards strategic goals. For many types of scientific projects (e.g., postdoctoral fellowships, EU ERC and NSF grants) it is very important who the principal investigator (PI) is. There are multiple definitions and understandings of PI's~\citep{Cunningham2016}. For instance, according to \citet{Melkers2012}, PI is a person, usually a senior researcher, who has won numerous grants and may assemble a scientific team to carry out the project under her scientific supervision. In our paper, PI is every researcher that was granted a project by the Slovenian Research Agency (ARRS)\footnote{\url{https://www.arrs.gov.si/en/index.asp}}; see Sections~\ref{sec:method} and~\ref{sec:basics} for more details.

In Slovenia, public research grants were delivered by the ministry responsible for science in the period 1994--2004, when ARRS was founded. Afterwards, all research projects have been granted through ARRS. We have retrieved complete data for \num{7508} projects out of \num{7576} research projects granted in the period 1994--2016. They have in total \num{2725} different PI's. We present more details and basic statistics about these projects in Section~\ref{sec:basics}.

The latest ARRS's call for research grants required from the PI of a public research project to fulfill various formal conditions. The quantitative evaluation of research projects includes as important part (e.g., $1/3$ of scores in the 1st round for basic research projects) the scientific excellence of a PI, which must be proven by outstanding publications (papers in highly ranked journals, books published with reputable publishers) and other research work (e.g., editorial work in high quality journals, invited lectures at important conferences, number of citations). To become the PI of a research project therefore outstanding previous research results must be demonstrated, cutting-edge results in the proposed project must be promised, and at the end a successful PI must fulfill all the commitments listed in the project proposal. For most researchers becoming a PI represents one of the most important landmark in their career~\citep{Cunningham2014, Cunningham2016}.

Becoming a PI therefore brings additional acceleration into PI's scientific career and naturally one may expect that PI's are outstanding researchers in terms of scientific excellence and therefore certify  that public money is well-spent. But is this really the case? Our literature overview shows that a very limited amount of research has been done on this topics (see Section \ref{sec:related_work} for a description of related work).

\subsection{Our contribution}

The main goal of this paper is to provide partial answer to the question posed in the title by comparing the scientific performance of
\begin{itemize}
\item [(i)] Slovenian researchers who were PI of at least one publicly funded research project; and
\item[(ii)] all active Slovenian researchers, i.e., those that were registered in Slovenian Current Research Information System (SiCRIS)\footnote{\url{http://www.sicris.si/public/jqm/cris.aspx?lang=eng&opdescr=home}} in June 2017 and have published at least one scientific publication in the period 1970--2016 (note that every PI is registered and there are only 4 PI's that have no scientific publication in this period). 
\end{itemize} 

We have chosen Slovenia because we have access to high quality data about researchers in Slovenia and about their research output.

The main contributions of this paper are:
\begin{itemize}
\item[(i)] We demonstrate that PI's have on average much better scientific performance compared to all active researchers.
\item[(ii)] We show that PI's have on average longer  and more fruitful career compared to all active researchers, with regards to all career indicators. In particular, PI's publication careers on average last approx.~$26.5$ years, while for all active researchers the average publication span lasts approx. $14.1$ years. 
\item[(iii)] The PI's that have received (among others) also a postdoctoral fellowship at the beginning of their career are at the beginning of their careers outperforming the other PI's. However, after the end of their postdoc project, their career productivity starts deviating towards the average career productivity of all PI's and $70\%$ of those who won postdoc project before 2011 (hence have finished it before 2014) did not succeed to win another research project by 2017.
\item [(iv)] On long run, the PI's leading the research programs (the most prestigious grants) on average demonstrate the best scientific performance.
\item[(v)] In the co-authorship network, the PI's form the backbone, i.e., removing PI's from the network results in a substantial fragmentation of the network. Additionally, we figured out that PI's create new PI's. More precisely, from 2004 on more than $90\%$ of researchers that become a PI for the first time, were connected in the co-authorship network with a PI.
\end{itemize} 

Our paper therefore highlights the relations between the public funding of the science and the scientific output for the case of Slovenia and contributes to better understanding how winning a public research grant and the scientific performance of a PI are related. The contributions (i), (ii) and (iv) also indicate that for the case of Slovenia the PI's are very important unofficial driver of science.

\section{Related work}
\label{sec:related_work}

In this section we first review important works on reseach community in Slovenia which are based on network analysis methodology. Then we highlight basic determinants of excellence in science. Finally, we conclude with state-of-the-art review of essential papers in the field of PI's.

In recent decades, we can observe a growing trend in exploration of different types of scholarly networks. In such networks a node usually refers to an author, a paper, or a journal, and relation between nodes is given by co-authorships, citations, or keywords. The most cited works on the analysis of large-scale bibliographic networks are authored by Newman and colleagues\footnote{\url{http://www-personal.umich.edu/~mejn}}. They described different universal characteristics of networks (e.g., small-worldness, assortative mixing, high modularity) which were later elaborated in many other studies.

In this paragraph we give general review of research work which was performed on Slovenian research community. The majority of material presented in these papers originates from the analysis of research evidence and statistics from two information sources, namely COBISS\footnote{\url{http://www.cobiss.si/cobiss_eng.html}} and SiCRIS, which are maintained by IZUM\footnote{\url{https://www.izum.si/en/}} and ARRS. \cite{Ferligoj2009} applied clustering of relational (and attribute) data to the collaboration network of Slovenian researchers and their publication performance. \cite{Perc2010a} studied the growth and structure of Slovenian research network for 1960--2010 concentrating mainly on verification of the small-world and preferential attachment assumptions. In addition, \cite{Perc2010b} showed that the distributions of citations per publication for various research fields and scientific institutions in Slovenia follow Zipf law. \cite{Kronegger2012} linked two distinct ways for describing co-authorship dynamics: small-world approach and preferential attachment. While they confirmed the presence of small-world structures, the preferential attachment mechanism seems considerably more complex and can not be attributed to specific global mechanism. Similar study was later conducted which used augmented bibliometric data for the complete longitudinal co-authorship network for all scientific disciplines~\cite{Ferligoj2015}. \cite{Luzar2014} developed an approach to study the community evolution in Slovenian's research network based on mutual collaborations. They reported that the interdisciplinary research collaboration is proportional to the overall growth of the network. \cite{Kronegger2015} studied classification of research disciplines including their longitudinal characteristics by concentrating on their collaboration structure over years. They found that clusters of disciplines for the Slovenian research network could be partitioned into five clusters that correspond to the official national classification of science. \cite{Karlovcec2015} investigated interdisciplinarity of research areas based on a network of collaboration between Slovenian researchers. Authors proposed novel statistic for interdisciplinarity that considers both network structure and content. \cite{KarLuzMla16} used machine learning approach to study the transition behaviour of researchers between core and periphery in the Slovenian research network. \cite{Hancean2016} studied how different network properties affect citation counts of sociologists of East European countries including Slovenia. Results indicated that the score of the citations of a researcher could be clearly predicted by the citation counts of her co-authors. Finally, \cite{Kastrin2016} addressed scientific creativity of Slovenian researchers from various perspectives which determine prolific science, including productivity, collaboration, internationality, and interdisciplinarity. They relate results to historical events and to domestic expenditure for research.

Let us now briefly review empirical evidence that deals with determinants of prolific science: productivity, collaboration, internationality, and interdisciplinarity. For deeper insight please consider~\citet{Kastrin2016}. We will use these determinants later in the paper to dissect Slovenian research community. First some words about productivity. It is well known that relatively small proportion of scientists write the majority of publications; this finding is known as Lotka's law~\citep{Lotka1926}. \cite{Pravdic1986} found that collaboration with prolific researchers generally increases personal productivity and \textit{vice versa}, collaboration with non-productive researchers decreases it. Collaboration among researchers is one of the most important characteristics of creative scientific work. Research collaboration studies were performed for different countries including Slovenia~\citep{Perc2010a}, Turkey~\citep{Cavusoglu2013}, Brazil~\citep{MenaChalco2014}, and Korea~\citep{Kim2016}. \citet{Lee2005} demonstrated that amount of academic papers is strongly correlated with the number of collaborators. However, researchers at the research institutes are more prone to collaborate than the researchers at the universities~\citep{Boardman2008}. Interdisciplinarity refers to integration of different scientific fields with the aim to create new research areas. \cite{Uzzi2013} pointed out that interdisciplinarity is of utmost importance for revolutionary scientific discoveries. Empirical evidence reveals that interdisciplinarity is more common in applied research~\citep{VanRijnsoever2011}. Internationality stimulates science to become greatly globalized. It was confirmed that internationality positively correlates with the quality of research~\citep{Katz1997}. Multinational papers are cited more frequently than papers from a single country~\citep{Glanzel2005}. \cite{Han2014} found that country-level collaboration is considerably mature, while international collaboration is still developing.

Current review of bibliographic databases reveals limited but growing body of literature on PI's. In the next paragraphs we summarise the most important findings of our review. We should emphasize the fact that a large body of reviewed literature refers to societal and psychological aspects of PI's, mainly from the perspective of leadership and research management.

There is a frequent belief that researchers can labor without leadership and that research could be directed without proper management quality. However, it is crucial to note that leadership is considered important to both researchers and research~\citep{Ball2007}. Proper leadership in scientific practice can boost research outcomes, organization eagerness, and commitment to research work~\citep{Bushaway2003,Moses1985}.

According to~\citet{Bland1992} a successful leadership is the most important characteristics that stimulates and maintains research productivity. When a researcher takes the PI responsibility she moderates her character from that of pure scientist to integrate much more other responsibilities and functions (e.g., forging goals, defining research projects, providing mentorship)~\citep{Jain2009}. \citet{Mangematin2014} argue that PI's are critical to the knowledge transformation and have become ``linchpins'' in shaping the scientific frontiers. \citet{Hemlin2006} found that effective leadership is more important to creative knowledge environments than organizational support. However, \citet{Boardman2014} addressed the question how PI's organize and manage researchers in scientific centers. Authors empirically demonstrated that some PI's exhibited managerial competences and some did not.

\citet{Casati2014} studied practices and activities of PI's to better understand their roles in science. Their main question was how PI's organize and coordinate research, how they handle different models of collaboration, and how they face with expanding complexity in science. Using in-depth interviews they derived four main practices of PI's: (i) focusing in research discipline, (ii) innovating and problem solving, (iii) shaping new theories and models, and (iv) brokering science. The former two practices tie strongly to project management, while the remaining two are very close to entrepreneurial activities. The involvement of PI's in different practices (e.g., brokering science) could be thus seen as a continua of activities. Similarly, \citet{Kane2015} categorized the strategic behaviours of PI's according to strategic position (proactive vs. reactive) and levels of funding. They described four classes of PI's, namely designer, adapter, supporter, and peruser.

\citet{Kidwell2014} performed qualitative methodological approach and identified three facets that illuminate the nature of PI's roles. First, PI's are visionars who mobilize numerous resources to enact their research agendas. Second, PI's are boundary spanning brokers in terms to expand their knowledge through discussions, engaging collaborations, and practicing good grantsmanship. Last but not least, PI's act as arbitrators and navigators among various tensions around them. Therefore, PI's employ various behavior strategies to pursue their research plan. For instance, PI's exhibit a propensity for welcoming new researchers in labs for finding new knowledge and opportunities for research collaborations. They are also very particular in selecting an institution where they run their research. Likewise, \citet{Baglieri2014} illustrated that some PI's make careful decision and set up their own company in order to establish new venture to enhance their impact and power.

\citet{Cunningham2014} seek to understand inhibiting factors that publicly funded PI's face. They identified three inhibiting factors: (i) political and environmental, (ii) institutional, and (iii) project based. Authors emphasize that these factors limit research autonomy of PI's. However, PI's have very small influence and control on these inhibiting factors although they have central role in publicly funded research. In the context of the Irish-based PI's \citet{Cunningham2015} identified three managerial challenges faced by PI's, namely project management, project adaptability, and project network management.

The commercialization of knowledge has become one of the key task within academic research~\citep{Miller2018}. Researchers thus also pay attention to the connection between academia and industry in which PI's have a leading role~\citep{Menter2016}. \citet{OReilly2017} focus on the observations of PI's and found that personal relationships, asset scarcity, and proximity issues act as barriers and enablers to technology transfer engagement with small- and medium-sized enterprises. In a similar vein \citet{OKane2017} studied the perspectives of PI's on the main inhibitors to commercialization through academic entrepreneurship. \citet{DelGiudice2017} explored the factors influencing the entrepreneurial attitudes of PI's and found out that country's culture could be a key element. Researchers also developed various theoretical models of PI's behavior~\citep{Leydesdorff1996}; for example, \citet{Cunningham2018} developed a micro level framework of PI's as value creators.

\citet{Feeney2014} deal with question, if PI's are more productive that co-PI's and those who do not have grants. They demonstrated being PI or co-PI is significantly correlated with research productivity; such researchers produce significantly more publications and supervise more PhD students. Additionally, PI's publish more than co-PI's. PI's also supervise more PhD students.

\section{Methodology}
\label{sec:method}

The main characteristics of our research is that we work with high quality data. We have retrieved a complete list of researchers that were PI's of public research grants delivered in the observed period from 1994 to 2016 and have related it with the high quality databases of all registered researchers in Slovenia (SiCRIS) and of (almost) all bibliographic production in Slovenia since 1970 (COBISS).

COBISS database started only in 1991, so it contains an almost complete production of Slovenian researchers (actually, also the production of non-researchers) after 1991, while the data about publications published before 1991 were entered later (after 1991) and probably lack some entries. However, we are not aware of any researcher, active in Slovenia in the last decade, that is not registered in SiCRIS or does not maintain her research production in COBISS. 

We consider all \emph{active} Slovenian researchers, i.e., the researchers that were registered in SiCRIS in June 2017 and have published at least one scientific publication\footnote{We say that a publication is \emph{scientific} if it is classified as an \emph{original scientific article}, a \emph{review article}, a \emph{short scientific article}, a \emph{published scientific conference contribution}, a \emph{published scientific conference contribution abstract}, an \emph{independent scientific component or a chapter in a monograph}, or a \emph{scientific monograph}} in the period 1970--2016 (we denote the set of all active researchers with $A$). Within $A$ we select a subset of all researchers that were PI's of at least one public research project since 1994 (we denote it by $\PI$). There were \num{19598} active researchers in this period and \num{2725} researchers out of them were PI's, i.e., $|A|=\num{19598}$ and $|\PI|=\num{2725}$. For both sets of researchers, we computed indicators about (career) productivity, (career) collaboration, (career) internationality and (career) interdisciplinarity.

We first recall some additional notation from \cite{Kastrin2016}. When we consider only the researchers from $A$ who published at least one publication in a year $y$, we denote this set by $A_y$ and call it the \emph{set of productive researchers in the year $y$}. Additionally, we denote with $P$ the set of all scientific publications published between 1970 and 2016 being co-authored by at least one registered researcher. Analogously as above, the subset of scientific publications published in a year $y$ is denoted by $P_y$ and $P_y(r)$ denotes the set of all publications from $P_y$ co-authored by the researcher $r$.

For every publication $p$ from $P$ published after 2003, when the CONOR database was introduced, we know the complete list of publication authors and hence we can count how many of them are registered in SiCRIS (for those we assume that they are local) and non-registered (i.e., international by our assumption). This has enabled us to calculate the researchers and publication internationalities.

All relevant data underlying this study are available from Zenodo repository\footnote{\url{https://doi.org/10.5281/zenodo.1211537}}.

\section{Overview of Slovenian publicly funded research projects}
\label{sec:basics}

In this section, we present basic statistics about research projects in Slovenia in the period from 1994 to 2016. In total, there have been \num{7576} research projects granted in the period 1994--2016. For $64$ projects we do not know the PI and for additional $4$ project the PI's have no scientific publication in the period 1970--2016. Therefore we focus in this paper on the remaining \num{7508} projects. They have in total \num{2725} different PI's, who have proposed the projects and have also taken responsibility to successfully accomplish it.  

For every project, we have the data about the year $y_s$ in which the project started and the year $y_e$ in which it ended. A project is \textit{active} in the year $y$ if $y_s \le y \le y_e$.

There are several standard types into which the projects granted by ARRS are classified. In what follows, we define all types of projects that are the subject of this study\footnote{\url{https://www.arrs.gov.si/en/akti/prav-sof-ocen-sprem-razisk-dej-sept-11.asp}} \footnote{\url{https://www.arrs.gov.si/en/analize/obseg01/crp.asp}} \footnote{\url{https://www.arrs.gov.si/en/medn/vodilna/}}:
\begin{itemize}
\item \textbf{research programme}: represents a comprehensive area of research work for which it is widely expected to be relevant and usable for an extended period and which has such an importance to Slovenia that it is considered as the national priority  according to the national R\&D programme. 
\item \textbf{basic project}: an original experimental or theoretical work, foremost aimed at obtaining new knowledge about the basis of phenomena and perceptible facts.
\item \textbf{applicative project}: an original research performed to acquire new knowledge. It is  directed primarily towards a practical goal or purpose. The applicative  project is not an industrial research or project in the field of experimental development.
\item \textbf{postdoctoral project}: a basic or applicative project performed by a single postdoctoral researcher who is also a PI of such project. A postdoctoral researcher is eligible to apply for such project if she has obtained  doctoral degree less than three years ago.
\item \textbf{infrastructure programme}: represents a maintenance of infrastructure as support to research in a public research institution or a research institution with concession in the form of instrumental support, support to science literature collections, popularisation of science and support to research programmes containing elements of an instrumental centre or science collection.
\item \textbf{targeted projects}: a mechanism which provides research support to the relevant ministries and other stakeholders, following the principle of ``evi\-dence-based decisions''.  Project themes are based on the proposals put forward by the ministries and other authorities responsible for specific areas of public interest.
\item \textbf{European projects}: (a) \textbf{Complementary Scheme} provides incentives for those applicants from Slovenian research organizations who were positively assessed by the ERC peer review process, and nevertheless not approved for co-financing. Applicants are invited to prepare an adjusted research project corresponding with main objectives of the original proposal. The objective of this scheme is to obtain favorable conditions for refining both researcher's scientific excellence and the original idea of the project. (b) \textbf{Lead Agency Procedure} promotes international scientific collaboration as a driving force for excellence. The cooperation agreement concluded by agencies from different countries enables researchers to apply for a joint project at one of the agencies (the Lead Agency). The application consists of a joint scientific description of the project, delineating the scientific contributions of the respective partners.
\end{itemize}

\begin{figure}[htp!]
\centering
\includegraphics[width=.75\textwidth]{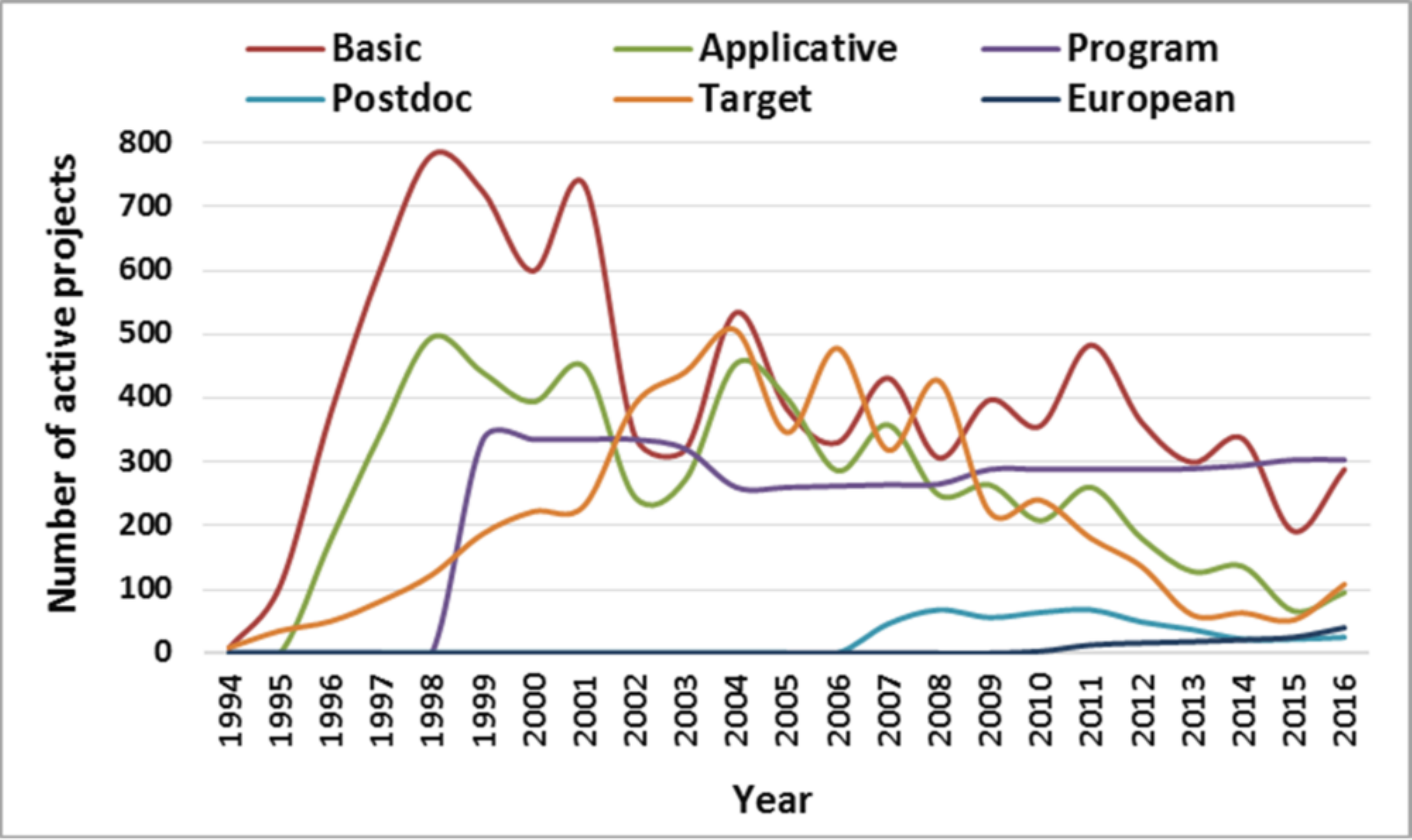}
\caption{The number of active projects in each year. The research programs are the most stable financing source of science.}
\label{fig:activeProjects}
\end{figure}

The most stable projects are the \textit{research programmes}, which last up to $6$ years and can be prolonged provided the goals are successfully acomplished. Such a schema provides continuous financing of research in a science field with, usually, allowing wider set of problems being considered, while projects of other types mostly focus in research of a very strictly defined topic. In Figure~\ref{fig:activeProjects}, it is shown that the number of active research programmes stayed the same during the last recession, while the number of all the others dropped dramatically.

Beside the research programmes, the most research oriented are the projects classified under the types \textit{basic}, \textit{applicative}, and \textit{postdoctoral}. These projects are usually granted in the same call. The postdoctoral projects are granted to researchers within a three-year period after obtaining PhD, to be able to initiate their own research. The principal investigators of these projects are the postdoctoral researchers themselves only after the year 2007, while before 2007, the PI's of such projects have been (usually) the mentors of the researchers. Thus, we classified postdoctoral projects as postdoctoral from 2007 on, and the earlier projects as basic projects. Additionally, for postdoctoral projects, in the set of the PI's we only take the researchers obtaining their project within a seven year period from their first publication.\footnote{We do this, since there are several researchers which published a scientific publication long before they obtained their PhD, and after PhD they also obtained a postdoctoral project, resulting in statistically irrelevant long tails in the career path analyses. We take seven years as a sum of four years of doctoral studies and three year period to obtain the project.}

\begin{figure}[htp!]
\centering
\begin{minipage}[c]{.48\textwidth}
\centering
\includegraphics[width=.95\textwidth]{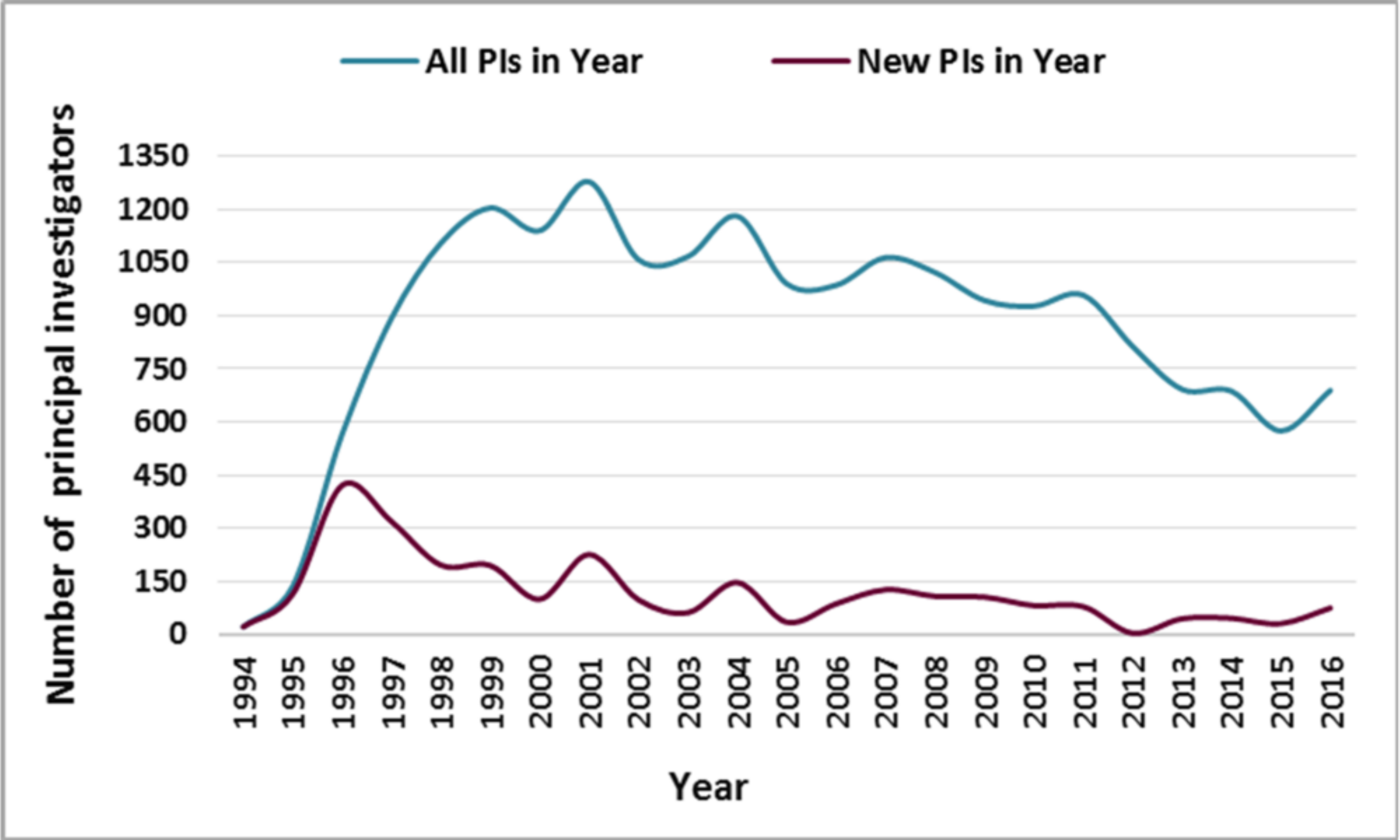}
\end{minipage}%
\begin{minipage}[c]{.48\textwidth}
\centering
\includegraphics[width=.95\textwidth]{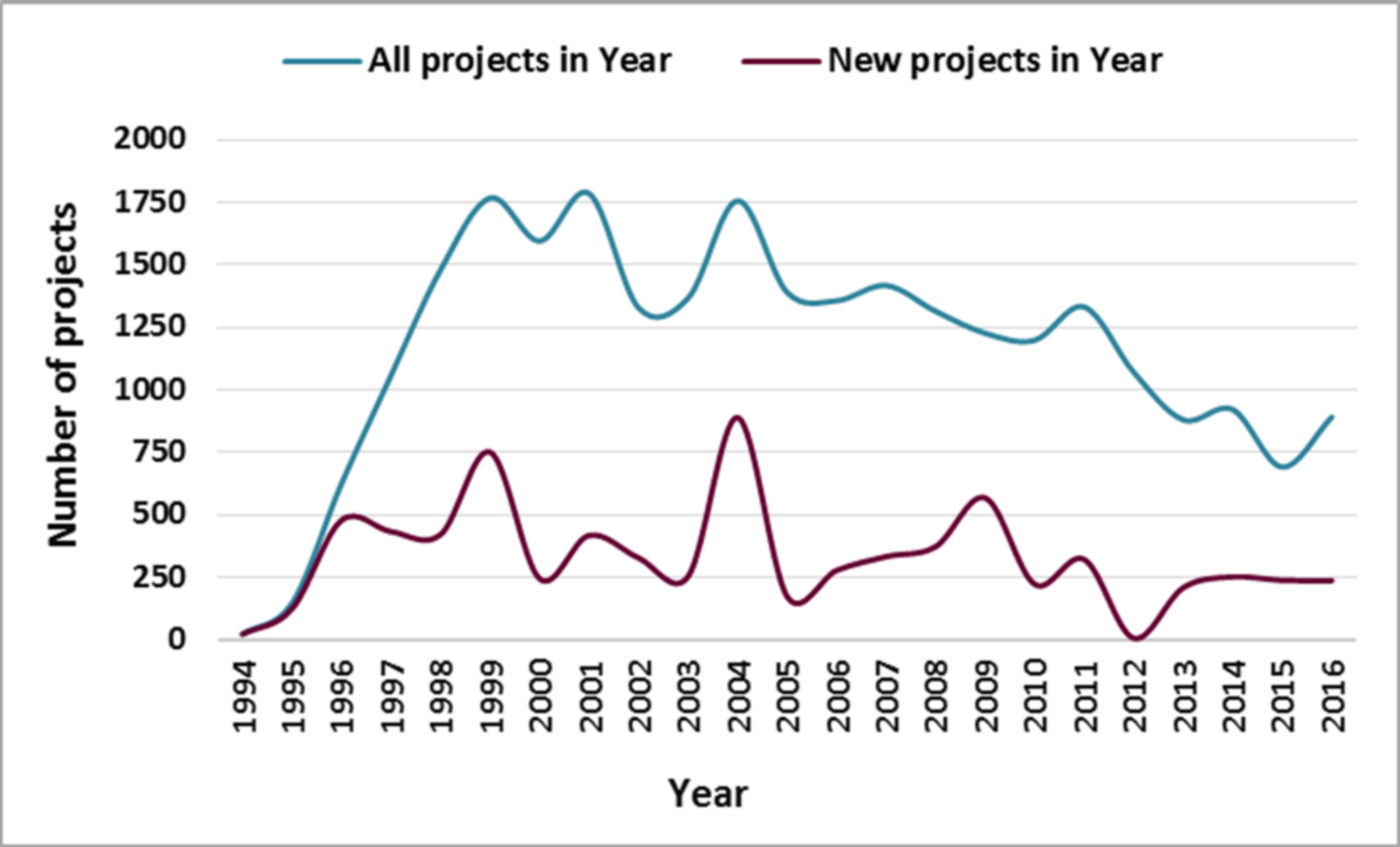}
\end{minipage}
\caption{The number of all and new principal investigators leading active project in given year (left). On the right we see the numbers of all projects active in given year and the number of projects that started in given year. }
\label{fig2:AllNewPIsAndProjects}
\end{figure}

In Figure \ref{fig2:AllNewPIsAndProjects} we depict the numbers of new and all PI's that were leading an active project in given year. Additionally we also show the number of new and all active projects for each year in time span of 1994--2016. We can observe that the numbers of all and of new projects were mostly decreasing since 2004. This is probably due to the fact that the budgets of the projects increased since 2004, while the number of projects decreased, especially in the period 2012--2015 (the public funding was stable until 2013 and then was decreasing until 2016 (see~\cite{Kastrin2016}, Figure~2). The year 2012 was extraordinary since in that year no new project started. Consequently, the numbers of all and of new PI's were decreasing too, with some fluctuations.

\section{PI's have outstanding scientific performance}
\label{sec:SciPerf}

In this section we report yearly results about productivity, collaboration, internationality, and interdisciplinarity of (i) PI's, i.e., the researchers from the set $\PI$, and of (ii) all active researchers, i.e., the researchers from the set $A$. These four indicators have been introduced and interpreted in \cite{Kastrin2016}. Recall, they consist of:
\begin{itemize}
\item \textbf{productivity}: average (fractional) number of scientific publications per researcher; 
\item \textbf{collaboration}: average number of (registered) collaborators for productive researchers (i.e., researchers who have published in given year at least one scientific publication); relative number of solo publications per year;
\item \textbf{internationality}: average publication internationality; average researcher internationality;
\item \textbf{interdisciplinarity}: average publication interdisciplinarity; average researcher interdisciplinarity.
\end{itemize}
They are not independent of each other, as it was demonstrated in \cite{Kastrin2016}.

\begin{figure}[htp!]
\centering
\includegraphics[width=.75\textwidth]{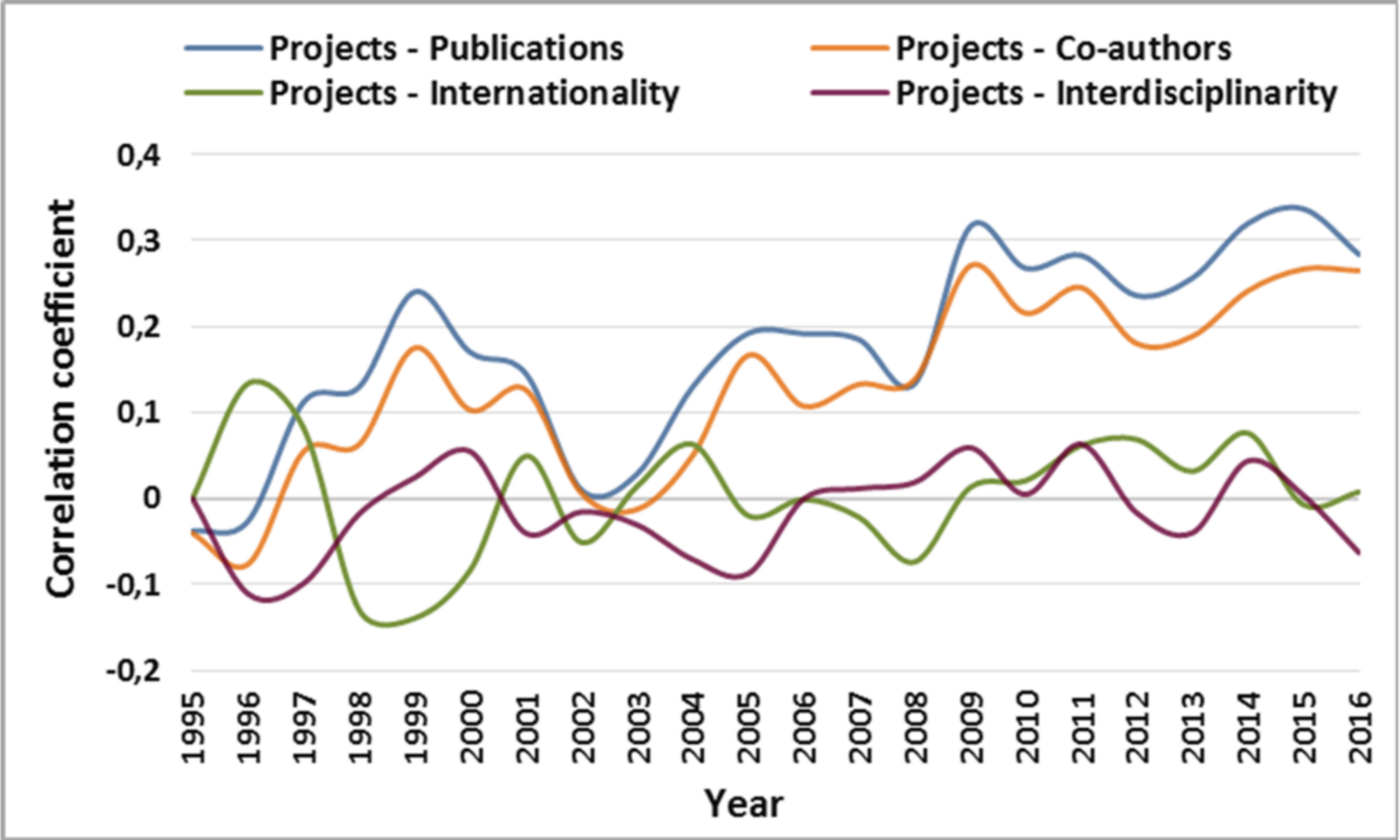}
\caption{The Pearson's correlation coefficients between the numbers of active projects in a given year and the annual values of average researchers' productivity, collaboration, internationality and interdisciplinarity. We can observe that the numbers of active projects are positively correlated with the annual values of average researchers productivity and collaboration, while correlations with annual values of average researchers internationality and interdisciplinarity are negligible.
}
\label{fig:correlation}
\end{figure}

We first show in Figure \ref{fig:correlation} how these indicators correlate with the number of active projects in a given year. The diagrams clearly show that the numbers of active projects are positively correlated with the annual average number of publications per productive researcher and with the annual average numbers of (registered) collaborators per productive researcher. On the other hand, the numbers of active projects are not correlated (or the correlations are very weak) with the annual average researcher's internationalities and annual average interdisciplinarities. 

We are not surprised with this, since large majority of projects are national and belong to one scientific domain. The only projects that by definition involve international teams are the European projects (described in Section~\ref{sec:basics}). In recent years, ARRS tries to stimulate interdisciplinary projects, but there are still only a handful of them.

\subsection{Productivity}
\label{subsec:productivity}

\begin{figure}[htp!]
\centering
\includegraphics[width=.75\textwidth]{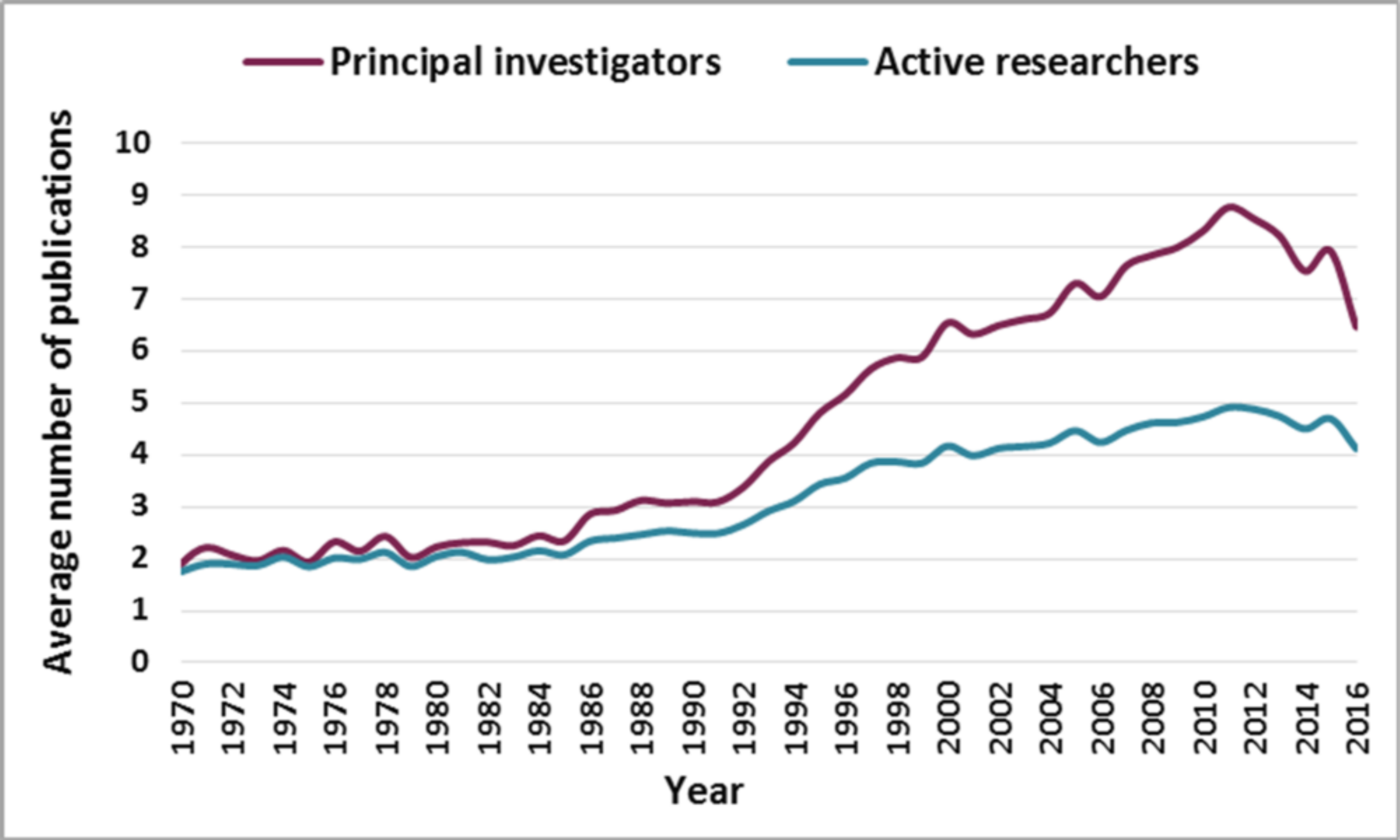}
\caption{The average number of publications per productive researcher per year shows is much larger for PI's than for all active researchers. This difference achieves the peak in 2011.}
\label{fig1:prod_pub}
\end{figure}

\begin{figure}[htp!]
\centering
\includegraphics[width=.75\textwidth]{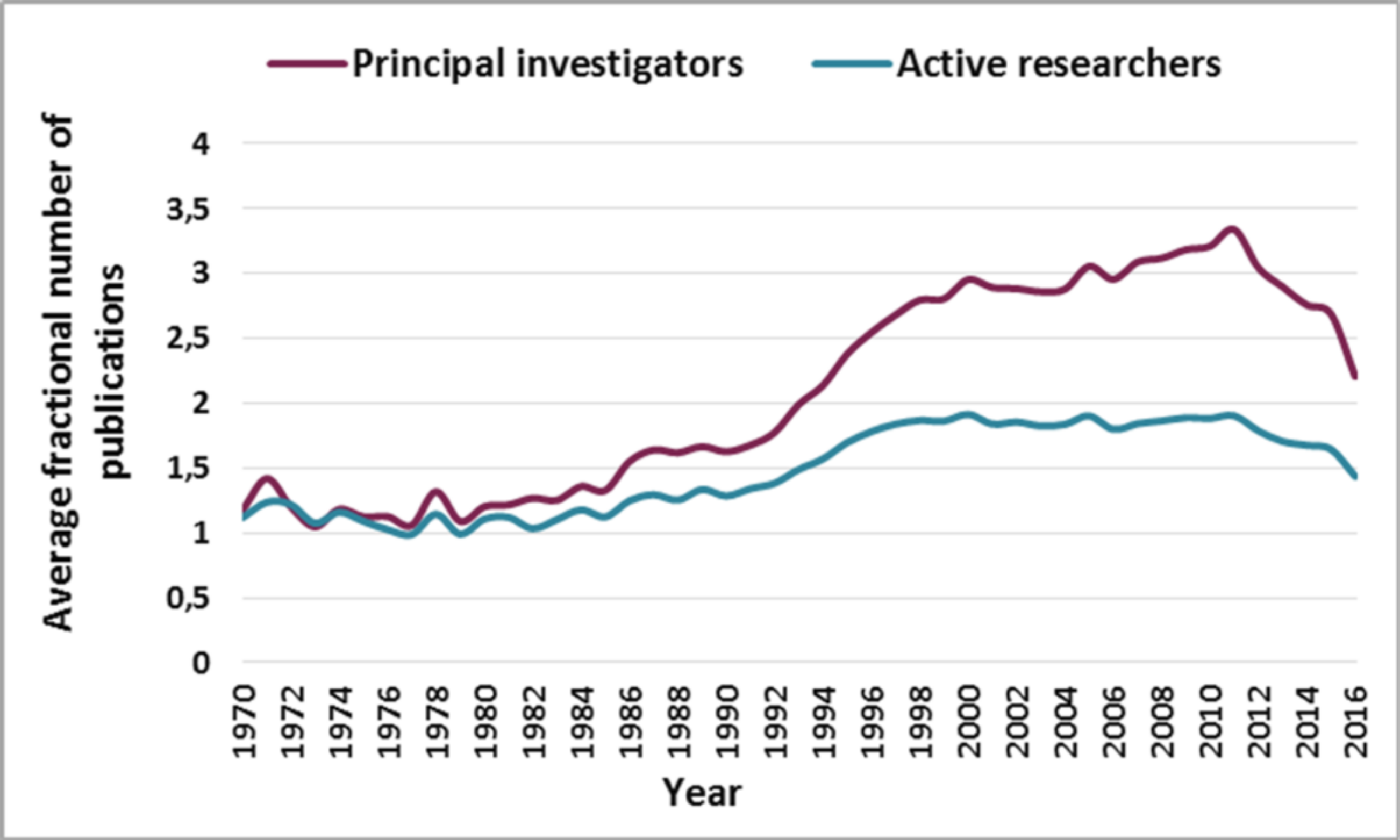}
\caption{The average fractional number of publications per productive researcher additionally demonstrates that a PI is on average scientifically more productive than an active researcher. The difference again achieves the peak in 2011.}
\label{fig2:prod_frac_pub}
\end{figure}

First, we calculate for both groups of researchers, $\PI$ and $A$, the average (fractional) number of publications per productive researcher for the period 1970--2016. The results are depicted in Figures~\ref{fig1:prod_pub} and \ref{fig2:prod_frac_pub}.

We can see that in terms of the average (fractional) number of publications per productive researcher the PI's are much more productive compared to all active researchers. Although the research grants that we consider started to be delivered only in 1994 one can notice that the PI's were slightly outperforming the others already in 1980s, i.e., a decade before they had become a PI.

However, a substantial difference can be observed after 1994 when the Slovenian research grants came into effect. The difference is largest in 2011 which was the year when governmental funding of research started to decline (the increase of funding has started again in 2016), see e.g. \cite[Figure 2]{Kastrin2016}. In 2011, the productive PI's have published on average approx.~$78\%$ more publications than (on average) all productive researchers. 

\subsection{Collaboration}

Collaboration of researchers from $\PI$ and $A$ was evaluated by calculating the annual average numbers of (registered) collaborators, co-authors of scientific publications, of productive researchers from $\PI$ and $A$, and the relative number of solo publications per year for researchers from $\PI$ and $A$ (again, we refer to \cite{Kastrin2016} for precise definitions).

\begin{figure}[htp!]
\centering
\includegraphics[width=.75\textwidth]{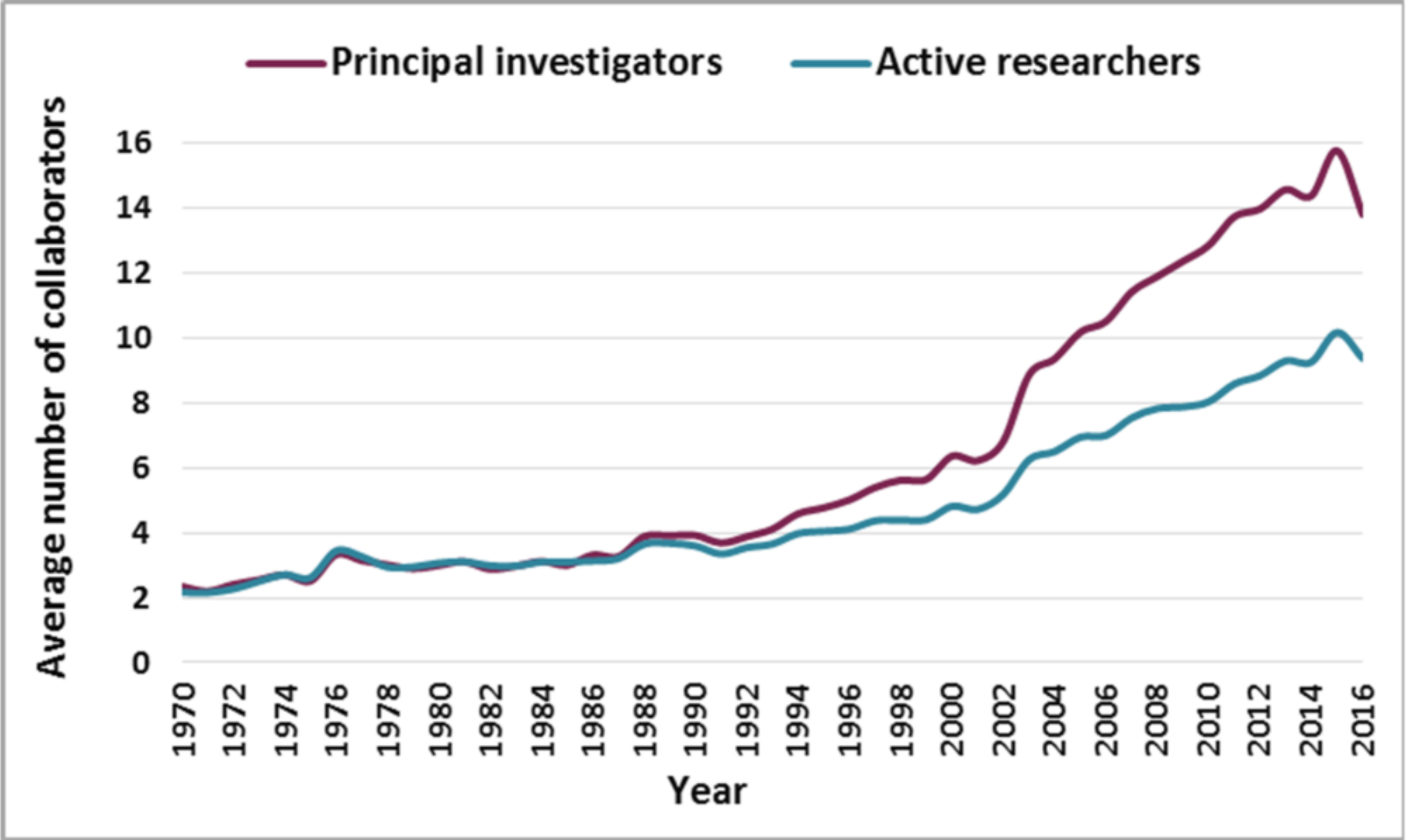}
\caption{The average number of collaborators per productive researcher per year for researchers from $\PI$ and $A$}
\label{fig3:prod_pub}
\end{figure}

\begin{figure}[htp!]
\centering
\includegraphics[width=.75\textwidth]{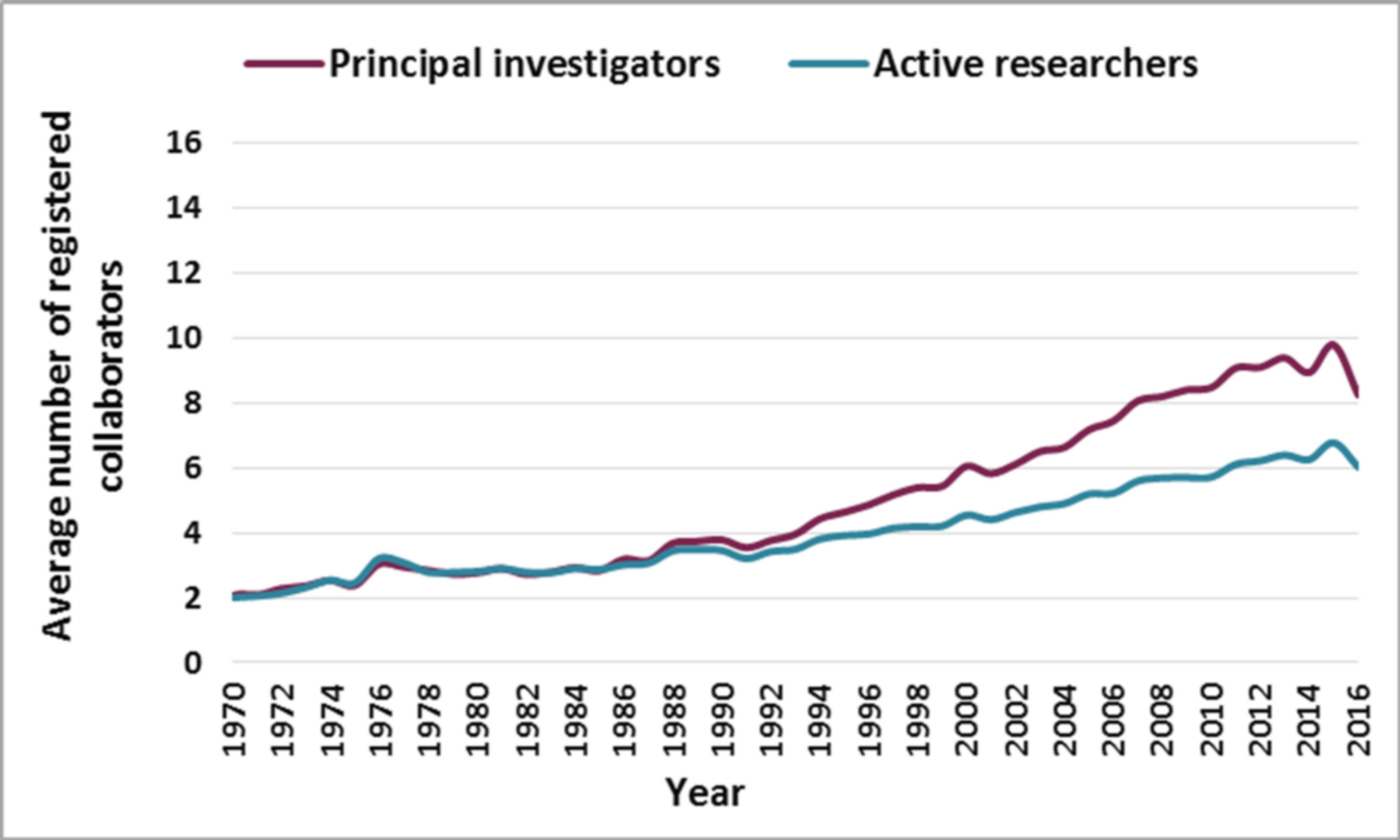}
\caption{The average number of registered collaborators per productive researcher per year for researchers from $\PI$ and $A$}
\label{fig4:prod_frac_pub}
\end{figure}

The annual average numbers of (registered) collaborators are depicted in Figures~\ref{fig3:prod_pub} and~\ref{fig4:prod_frac_pub}. The difference between these two indicators is that the second involves only the collaborators that are members of $A$, while the first indicator takes into account all co-authors of scientific publications, e.g. beside those from $A$ also non-registered Slovenian researchers (there are only few of them) and the foreign collaborators.

We can observe in Figures~\ref{fig3:prod_pub} and~\ref{fig4:prod_frac_pub} that PI's have on average much more co-authors on their scientific publications. The largest difference is again in 2011 when a productive PI has on average published her scientific publications with approx.~$59\%$ ($47\%$) more (registered) co-authors compared to all active researchers in this year. If we compare collaboration and productivity, we can observe an interesting fact: the average numbers of publications and of collaborators per productive researcher are increasing, see Figures~\ref{fig1:prod_pub}, \ref{fig3:prod_pub}, and \ref{fig4:prod_frac_pub}. But the average fractional number of publications per productive researcher is much more stable, especially for all active researchers. This means that the increase in the number of publications can be explained by more collaboration (i.e., on average more co-authors of publications), which is related to the \textit{publish-or-perish} effect. Usually, it is more appreciated (and also scored) if a researcher publishes more papers as a co-author than few papers alone or with one co-author. 

\begin{figure}[htp!]
\centering
\includegraphics[width=.75\textwidth]{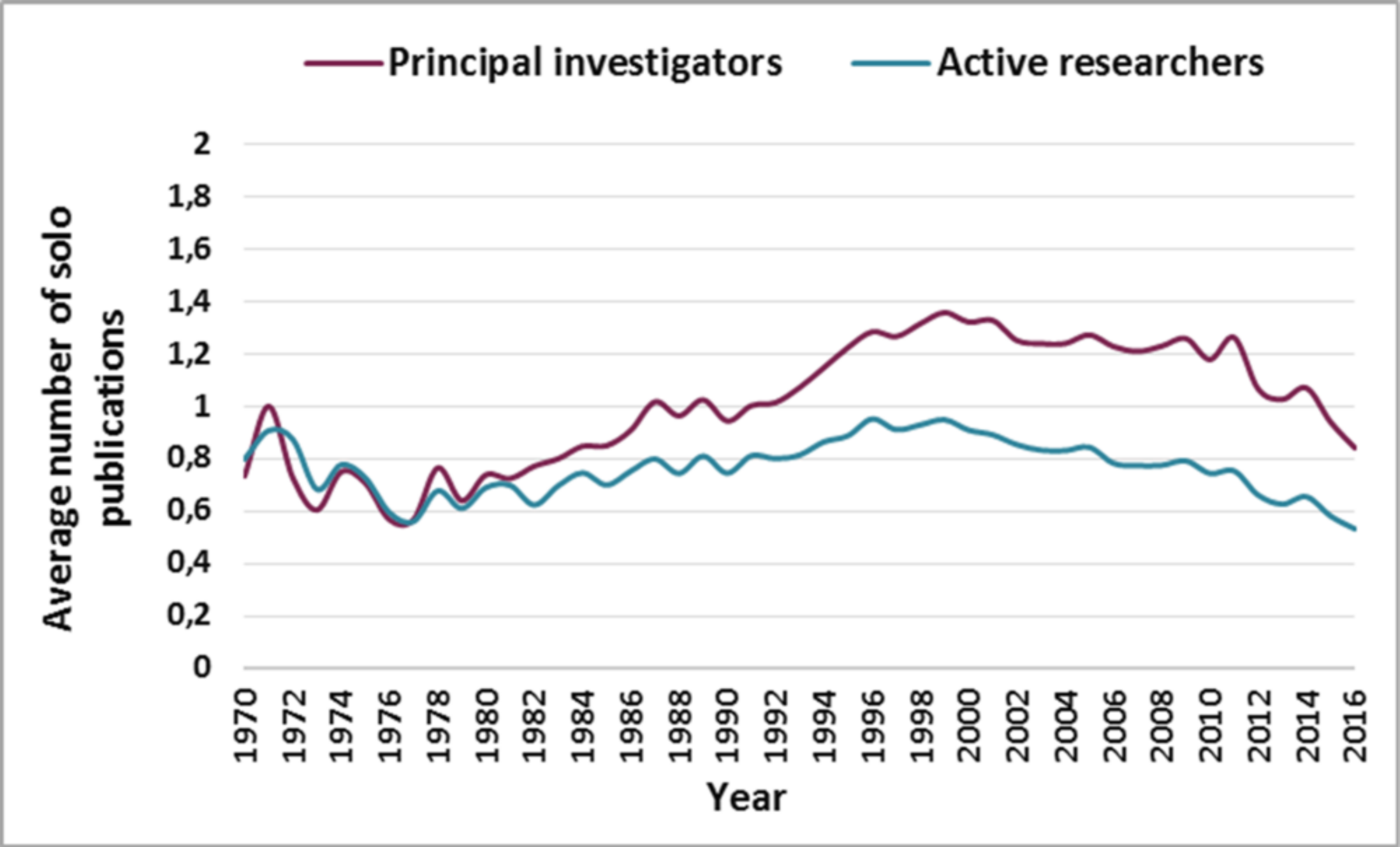}
\caption{The average number of solo publications per year for researchers from $\PI$ and $A$. PI's have on average more solo publications, but in general, these average numbers are decreasing since 2000.}
\label{fig5:soloPubs}
\end{figure}

Figure~\ref{fig5:soloPubs} depicts that PI's have on average much higher number of solo publications (scientific publications with only one author) compared to all active researchers, which can be easily explained by our assumption that PI's have much better scientific performance and that should be visible also in publishing without co-authors. However, it is evident that the average number is decreasing since the beginning of the century, while the average number of all publications is increasing. We see the main reason again in the \textit{publish-or-perish} trend.

\subsection{Internationality}

\begin{figure}[htp!]
\centering
\includegraphics[width=.75\textwidth]{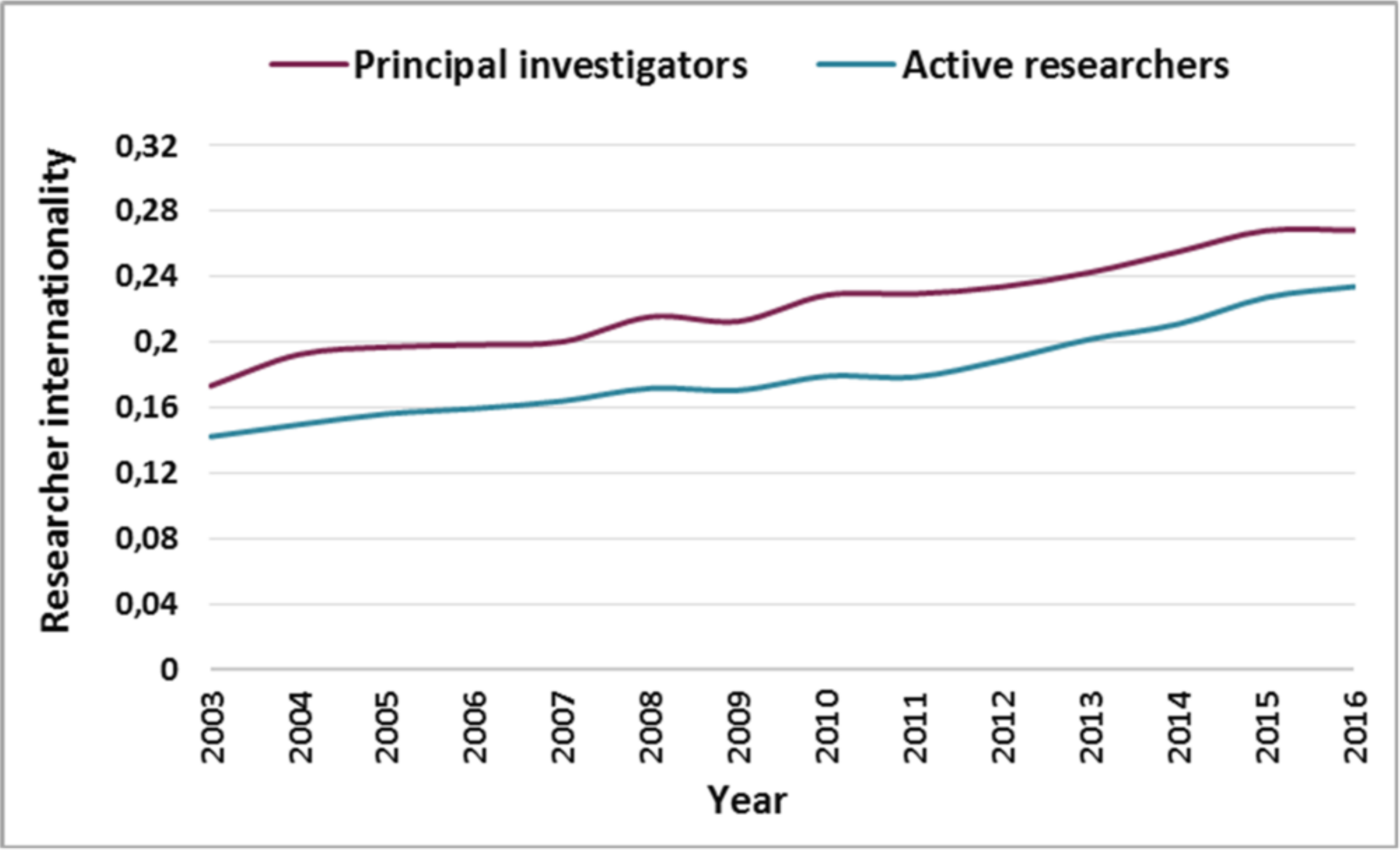}
\caption{The average researcher internationalities of researchers from $\PI$ is higher than the average researcher internationalities for $A$}
\label{fig6:InternatRes}
\end{figure}

Researcher's internationality of a given researcher for a given year is defined as the quotient between the number of all international co-authors of this researcher in this year and the number of all co-authors of this researcher in this year. We average these numbers over $\PI$ and $A$. Figure~\ref{fig6:InternatRes} depicts how both average researcher internationalities vary in the period 2003--2016. Recall that we are able to count registered and non-registered (i.e., international by our assumption) only from 2003, due to deployment of CONOR (see Section~\ref{sec:method}). We can see that PI's have on average significantly higher annual internationalities compared to all active researchers. Although yearly numbers of internationalities and active projects are not correlated, as demonstrated in Figure \ref{fig:correlation}, the average annual internationalities of PI's are higher. This is not surprising, since PI's manage project budgets and can therefore afford much more travelling, hence can easier create new international collaborations.

\begin{figure}[htp!]
\centering
\includegraphics[width=.75\textwidth]{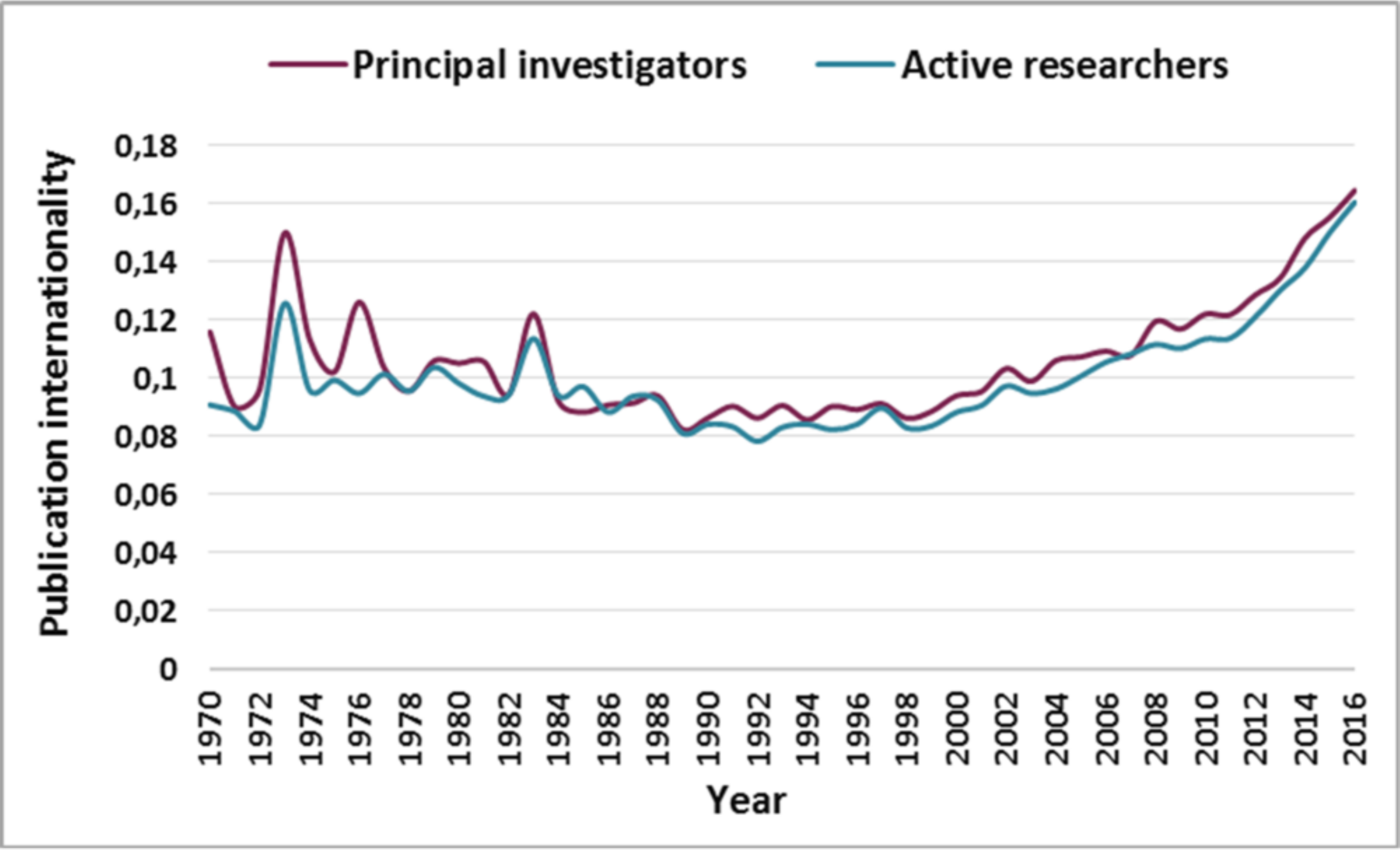}
\caption{The average publication internationalities for researchers from $\PI$ and $A$ increase from 1994 on. PI's on average  slightly outperform all  active researchers.}
\label{fig7:InternatPubs}
\end{figure}

Similarly to researcher's internationality, we define and calculate the average publication internationality---here we take for each publication the proportion of international authors, see Figure \ref{fig7:InternatPubs}. Unlike researcher internationalities, we are able to compute publication internationalities since 1970 as for every scientific publication we have the number of all co-authors and a list of the co-authors registered in SiCRIS. Recall that we consider the co-author as international if she is not in $A$. This is actually not completely correct since Slovenian researchers that moved abroad are often still registered and hence considered ``national''. Likewise the international (foreign) researchers who moved to Slovenia are almost in all cases registered in SiCRIS, hence considered ``national''. We have checked several names manually to convince ourselves that this happens very rarely, but we do not have data that would enable us to quantify this phenomenon.

Additionally, sometimes a researcher or a developer from (local) industry appears as co-author. If she is not registered we consider her as ``international''. We are convinced that these deviation also appears rarely. Indeed, there are some indirect indicators that support our claim. For example, you can not apply or work on a research grant funded by Slovenian Research Agency if you are not registered in SiCRIS. Even more, almost every funding instrument for basic or applied (including industrial) research that is based on Slovenian public money demands that the researchers involved in the project are registered in SiCRIS.

We neglect these deviations and consider all non-registered researchers as international.

Figure~\ref{fig7:InternatPubs} shows that average annual publication internationalities of PIs' publications do not differ significantly compared to average annual publication internationality of all active researchers. This actually means that PI's publish more and have larger international research networks but on each particular publication the proportion of international co-authors is not outstanding. This is probably aligned with the fact that almost all granted ARRS projects (more than $99\%$) were given to national consortia---only in the last few years international projects are also possible (see Section \ref{sec:basics}, item European projects - Lead Agency Procedure) and by the end of 2016 there were only $60$ such projects. Last but not least, internationality remains very complex problem which could not be described using only bibliometric attributes~\cite{Glanzel2001,Glanzel2005}.

\subsection{Interdisciplinarity}

Researcher (publication) interdisciplinarity measures from how many different scientific fields the co-authors of a given researcher (authors of a scientific publication) in a given year are coming. The formal definitions are rather involved so we again refer to~\citet[Subsection~4.4]{Kastrin2016}. We only recall that for each active researcher we get from SiCRIS database only information about her current scientific field (as valid in June 2017). The changes of scientific fields usually happen when researcher changes the job (moves from one research group to another) but mostly they happen on the second or third level of a scientific field (hence within one of the six main fields we are considering).

\begin{figure}[htp!]
\centering
\includegraphics[width=.75\textwidth]{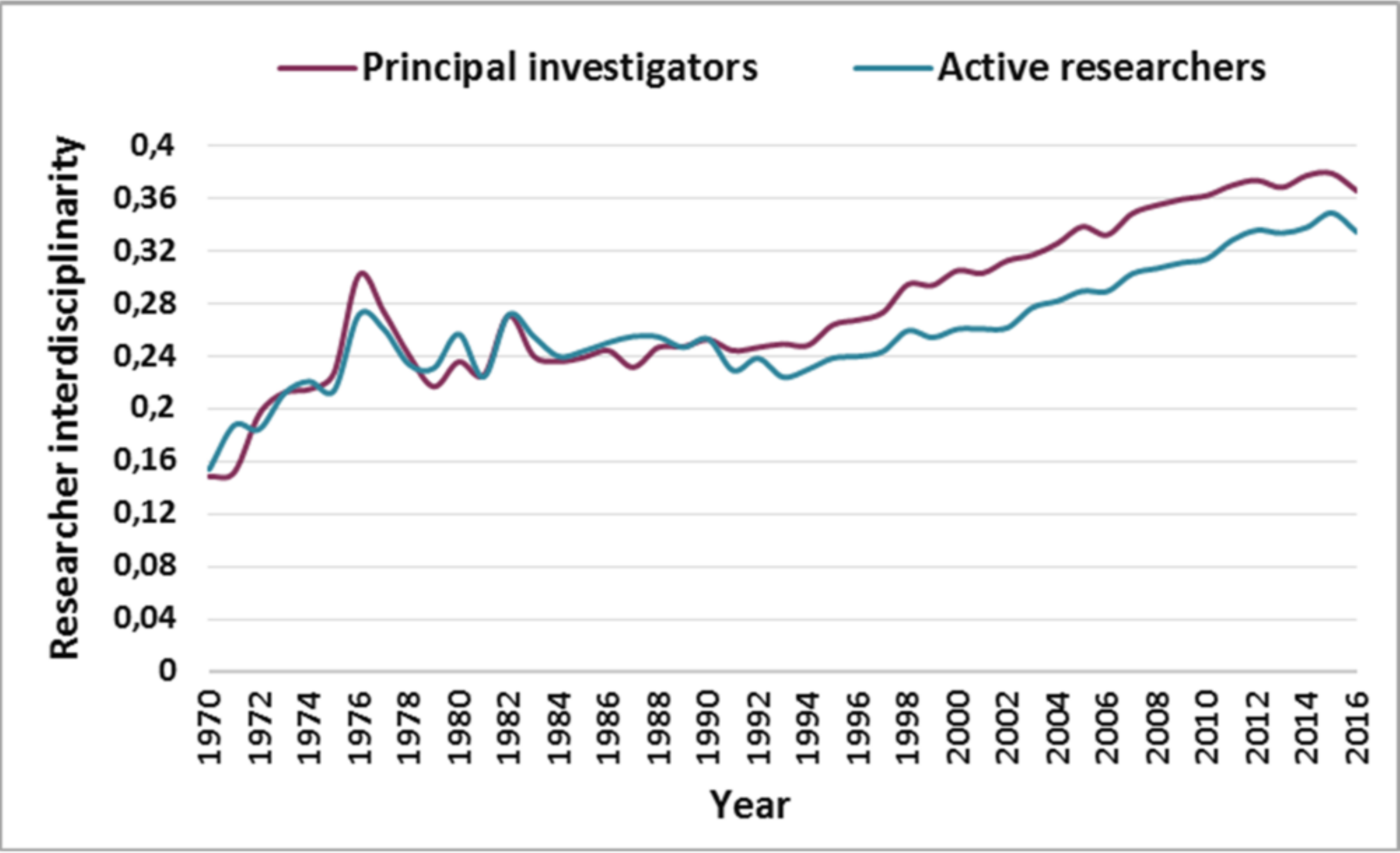}
\caption{The average researcher interdisciplinarities for researchers from $\PI$ and $A$ are increasing in the period that we mainly observe (1994--2016). PI's are on average in this period more interdisciplinary compared to all active researchers.}
\label{fig8:InterdiscRes}
\end{figure}

\begin{figure}[htp!]
\centering
\includegraphics[width=.75\textwidth]{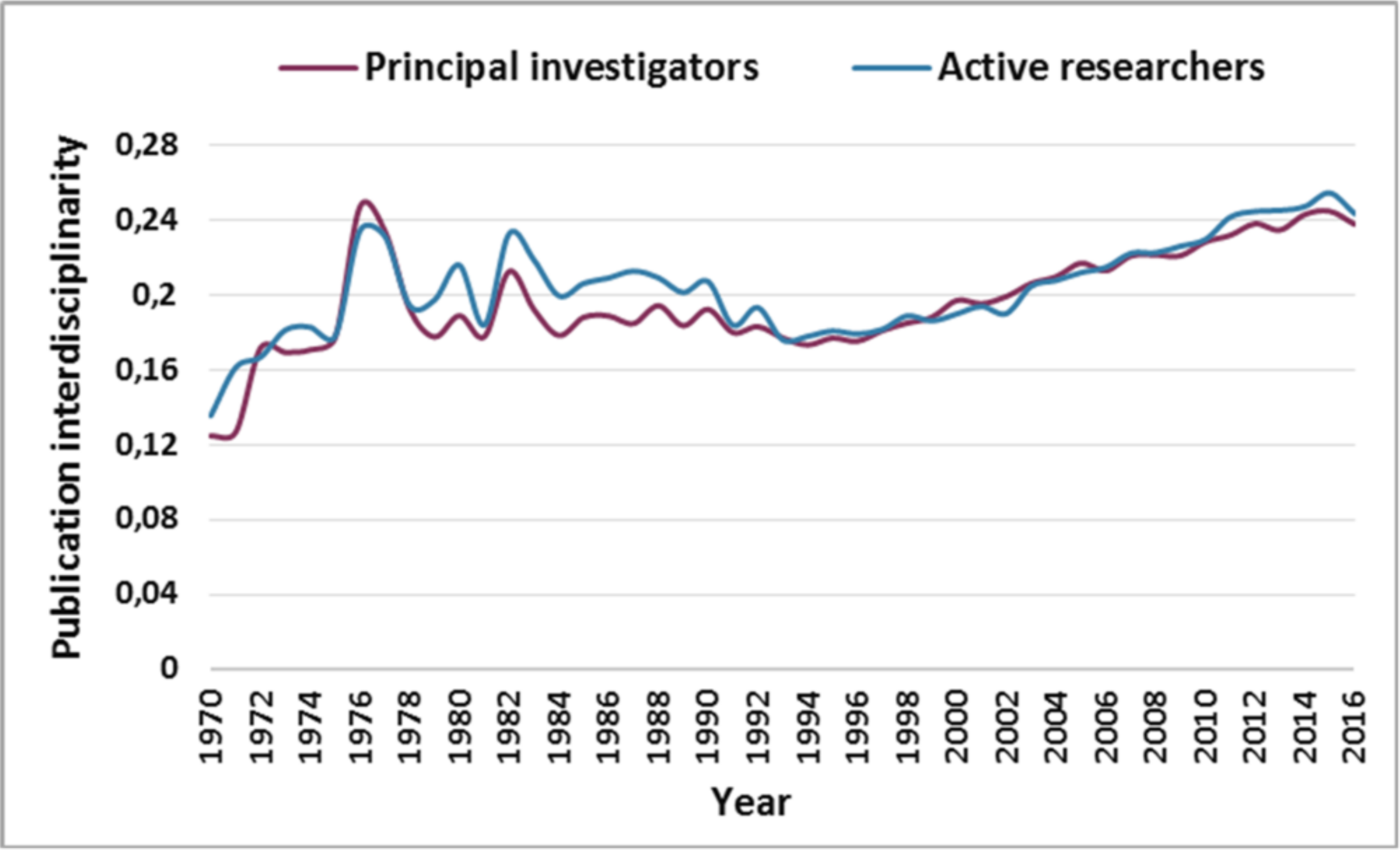}
\caption{The average publication interdisciplinarities for researchers from $\PI$ and $A$ are increasing in the period that we mainly focus to (1994--2016) and are very similar to each other}
\label{fig9:InterdiscPubs}
\end{figure}

We depict annual researcher's and publication's interdisciplinarities, calculated for PI's and for all active researchers in Figures~\ref{fig8:InterdiscRes} and~\ref{fig9:InterdiscPubs}. Figure~\ref{fig8:InterdiscRes} demonstrates that PI's have after the beginning of 1990s slightly larger researcher's interdisciplinarity compared to all active researchers. On the other hand, the publication interdisciplinarities for publications of PI's and of all active researchers do not differ significantly. This means that PI's are working in more research teams having members from different scientific areas but each particular (sub)project and publication based on it was mainly done by a research group with an average interdisciplinarity.

\section{PI's have on average more fruitful career}

In this section we present results about career performance of PI's compared to career performance of all active researchers. For that, we use a variation of four determinants investigated in the previous section. We consider the \textbf{career productivity}, introduced in \cite{Kastrin2016}, for both groups of researchers. We have improved the definition of this indicator such that the publication results obtained in different periods are now more comparable. In addition, we have extended this concept to \textbf{career collaboration}, \textbf{career internationality} and \textbf{career interdisciplinarity}.

Following \cite{Kastrin2016}, for each researcher $r\in A$ we have determined the year of the first publication, denoted by $\st{r}$, and the year of the last publication, denoted by $\en{r}$. The $r$'s \emph{publication career} therefore spans over the publication career years (\PCY) from $1$ to $\en{r}-\st{r}+1$. 

\begin{figure}[htp!]
\centering
\includegraphics[width=.75\textwidth]{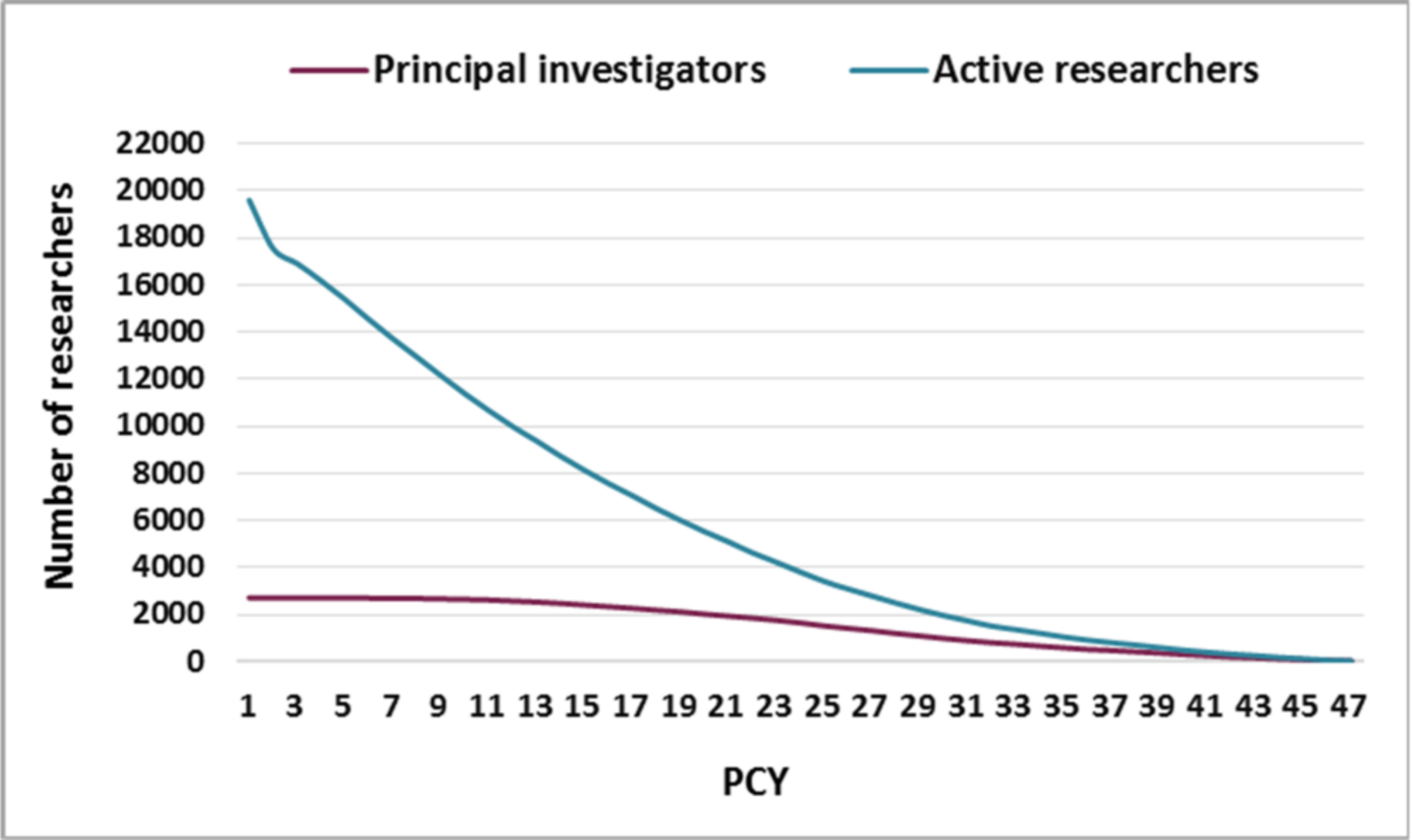}
\caption{The numbers of active researchers and PI's  for $\PCY=1,2,\ldots,47$. The number of PI's is decreasing slowly as on average their academic career is much longer. Huge gap at the beginning is mainly caused by the active researchers who finish their academic career after PhD, which is soon after their first publication year.}
\label{fig10:PCY_num}
\end{figure}

For each $\PCY$, we compute the number of PI's and the number of active researchers that were still active in this $\PCY$ (note that researcher $r$ is still active in \PCY ~if~$\PCY \le \en{r}-\st{r}+1$), see Figure~\ref{fig10:PCY_num}.

We can see that the population of PI's is stable and decreases much slower compared to the population of all active researchers. PI's academic career lasts on average $26.5$ years, while this average is for all active researcher only $14.1$ years.\footnote{Such a big difference surprised us, so we computed these averages also for both groups after ignoring (i) the researchers that have stopped within 3 years after the first publication (they probably had left academia after PhD, so are not typical members of the research community) and (ii) the researchers that are still active, since we do not know when they will stop (we kept only those who haven't publish anything in the years 2015--2016; they probably had stopped their publication career already). However, the gap remains almost the same (the average for PI's was $26.4$ and the average for all active researchers was $15.3$). } 

\begin{figure}[htp!]
\centering
\includegraphics[width=.75\textwidth]{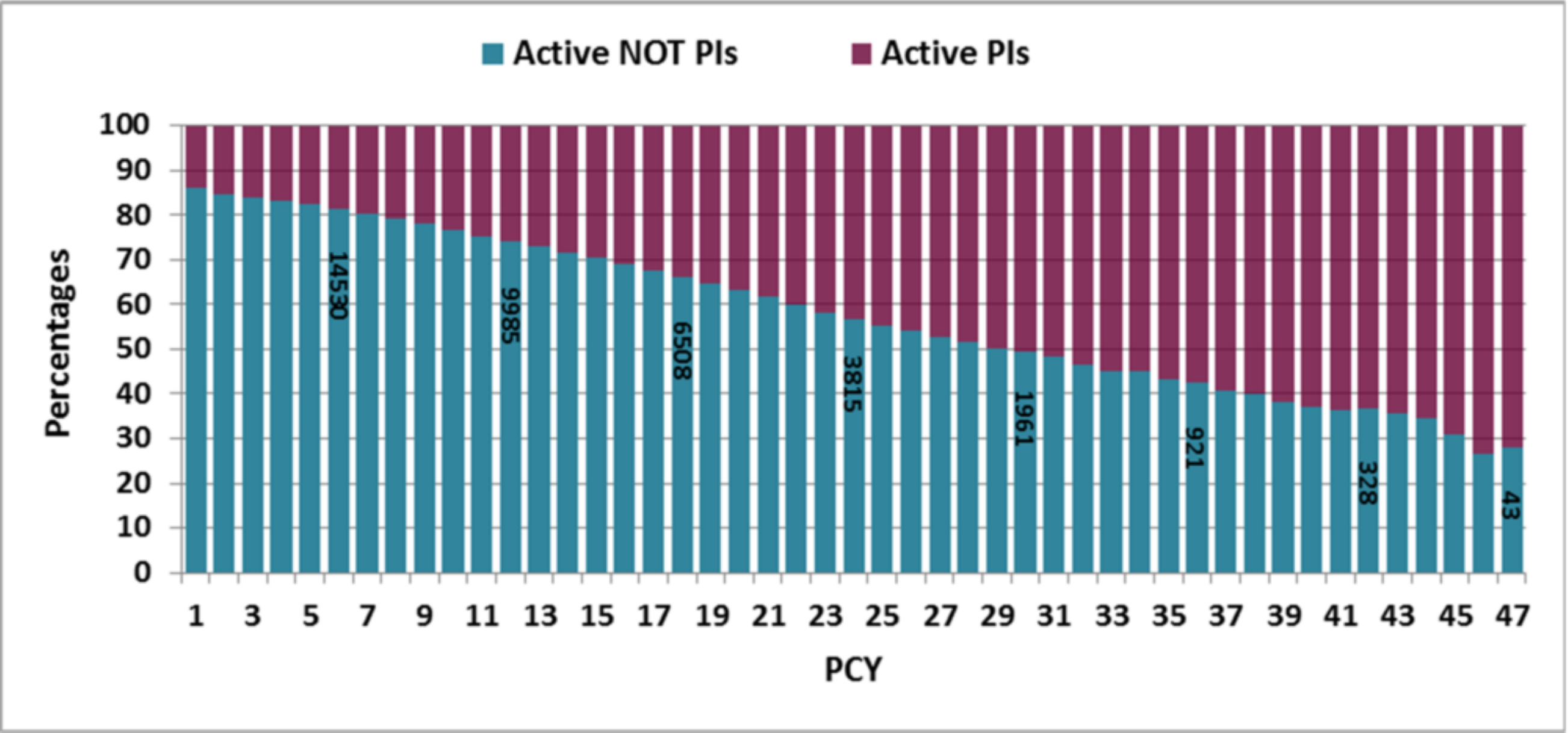}
\caption{The percentage of PI's among active researchers in PCYs from $1$ to $47$. For each sixth year we provide also the number of all active researchers. One can see that the majority of the researchers having long careers are PI's.}
\label{fig11:PCY_proportion}
\end{figure}

Figure~\ref{fig11:PCY_proportion} depicts that the researchers having a long and productive career are mostly the PI's---more than $50\%$ of researchers that are active more than $30$ years are PI's. This percentages increase to $63\%$ after the year $40$.

The set of researchers that are active in given PCY is very diverse. For example, for $\PCY=1$ such set contains all researchers that have published at least one scientific publication in their career, i.e., it is exactly $A$ and therefore includes also the researchers with the first publication in 1970 and the researchers with the first publication in 2016. In Subsection \ref{subsec:productivity} we show that the productivity in terms of average number of publications has increased a lot since 1970, so computing the average number of publications per researcher within the same PCY means that we align the productivity of researchers that has just started career with the productivity of researchers that are at the end of their career, which is not appropriate. 
 
We decided to eliminate this drawback by computing for each active researcher for each calendar year the normalised productivity, which is the quotient between the number of her publications in this year and the average number of publications in the same year, taken over all productive researchers. 

We call such productivity \emph{normalised productivity} of given researcher in observed (calendar) year. Finally, we compute for every PCY the average of all normalised productivities of all researchers active in this PCY. We call this indicator \emph{career productivity}.
  
Similarly we introduce and compute \emph{career collaboration}, \emph{career internationality}, and \emph{career interdisciplinarity} by computing respectively the average normalised number of registered co-authors, the average normalised researcher's internationality and the average normalised researcher's interdisciplinarity for $\PCY=1,\ldots,47$ for all members of $\PI$ and $A$ that were  active in given $\PCY$.

\begin{figure}
\centering
\begin{minipage}[c]{.48\textwidth}
\centering
\includegraphics[width=.95\textwidth]{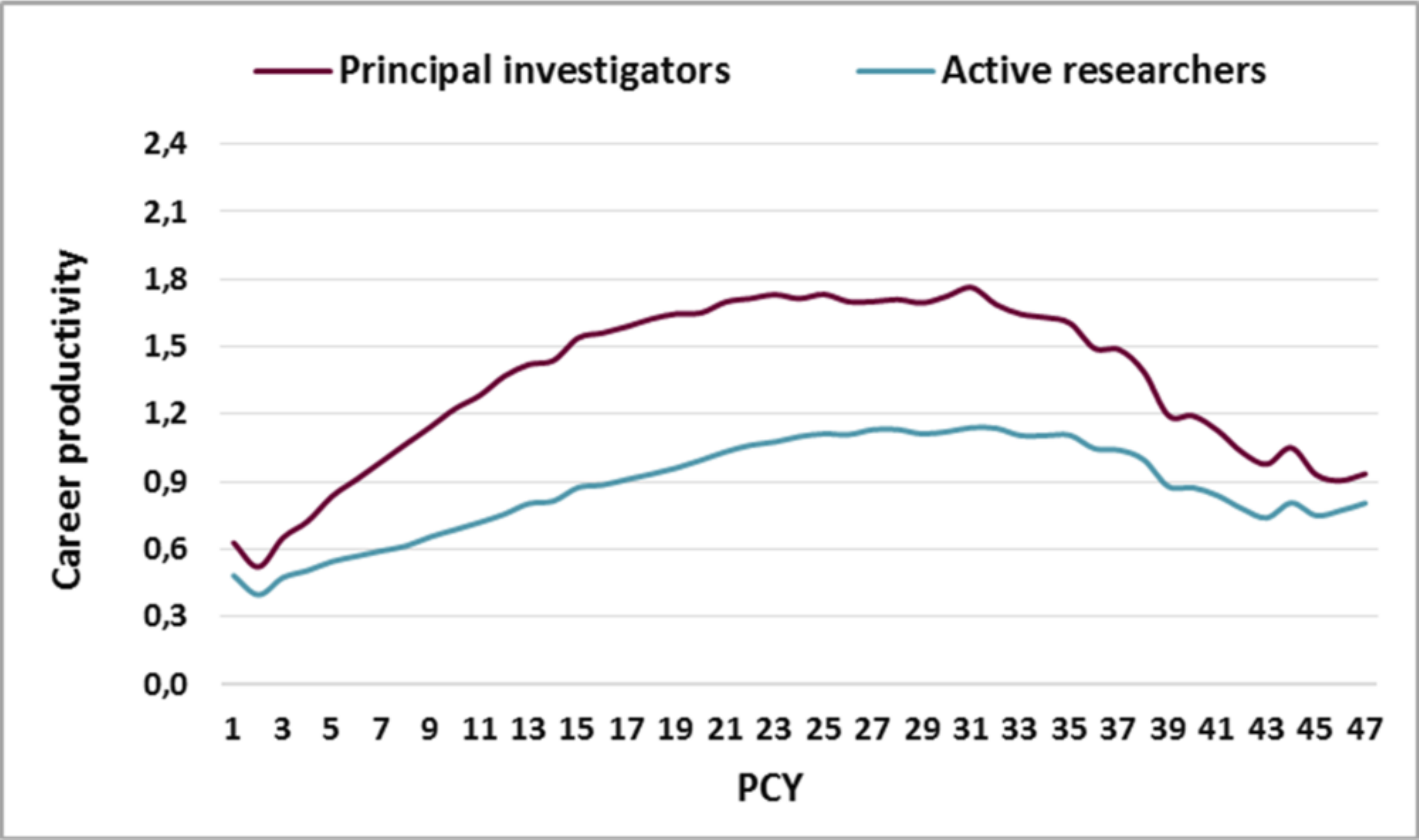}
\end{minipage}%
\begin{minipage}[c]{.48\textwidth}
\centering
\includegraphics[width=.95\textwidth]{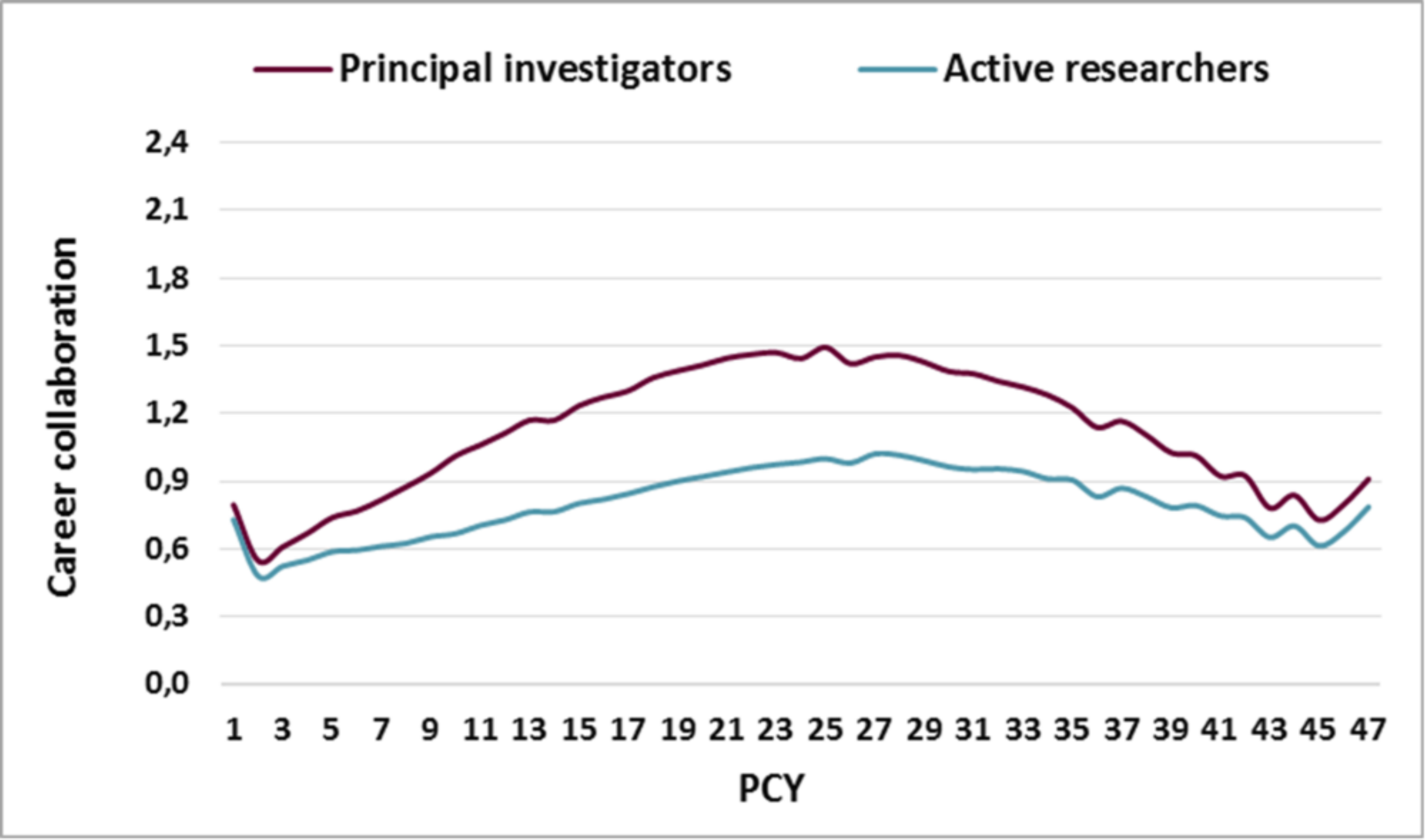}
\end{minipage}
\caption{Career productivity (left) and collaboration (right) for PI's and for all active researchers. We can see that PI's have much more productive and collaborative careers. In both graphs there is a drop in the second year, due to the fact that we consider researcher's career since her first publication year. In many cases, the researcher does not publish anything in the second year.}
\label{fig12:CareerProductInternat}
\end{figure}

Figure~\ref{fig12:CareerProductInternat} depicts the dynamics of career productivity and collaboration. We can observe that the academic career of PI is on average much more productive and collaborative, which is consistent with observations from Section~\ref{sec:SciPerf}.

\begin{figure}
\centering
\begin{minipage}[c]{.48\textwidth}
\centering
\includegraphics[width=.95\textwidth]{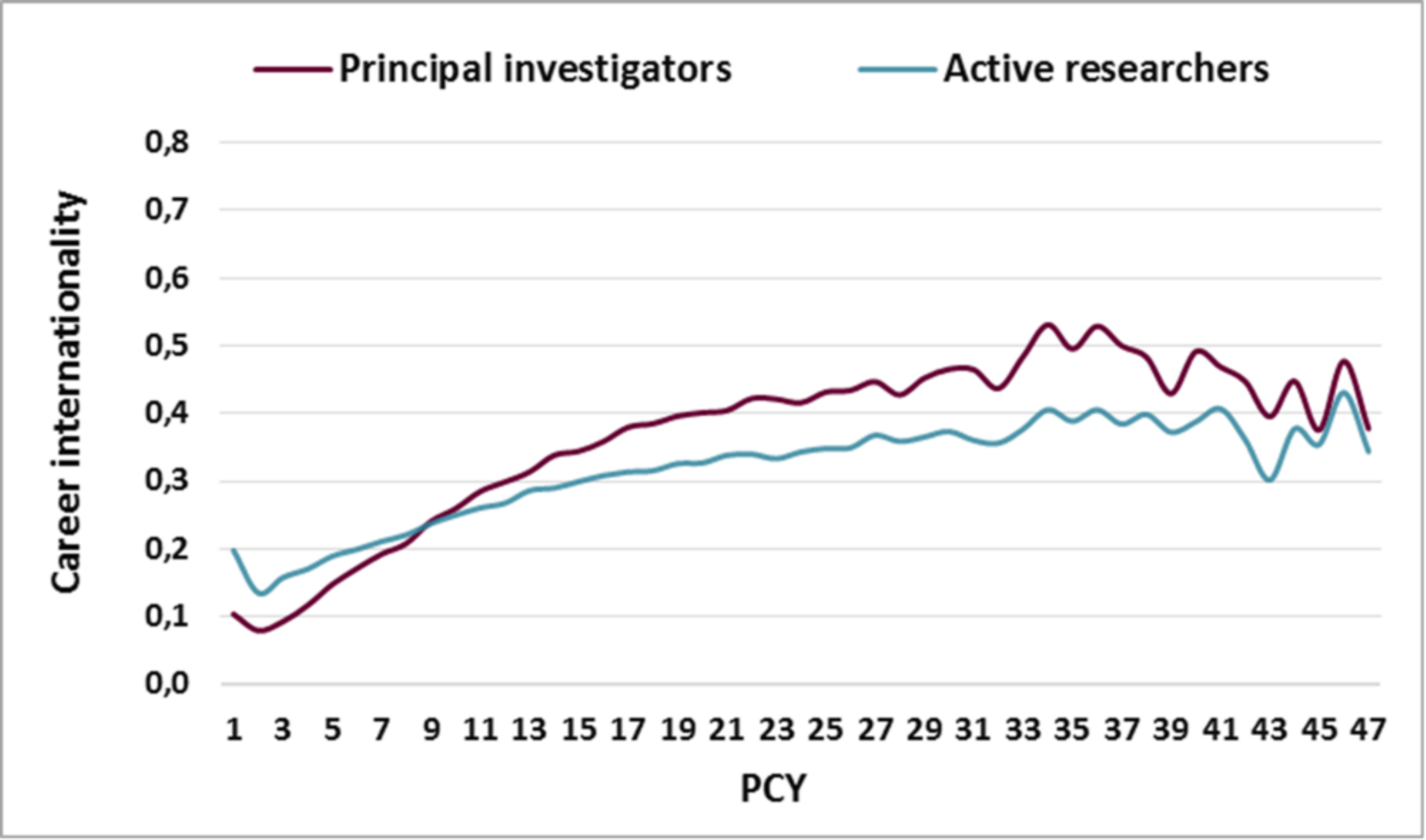}
\end{minipage}%
\begin{minipage}[c]{.48\textwidth}
\centering
\includegraphics[width=.95\textwidth]{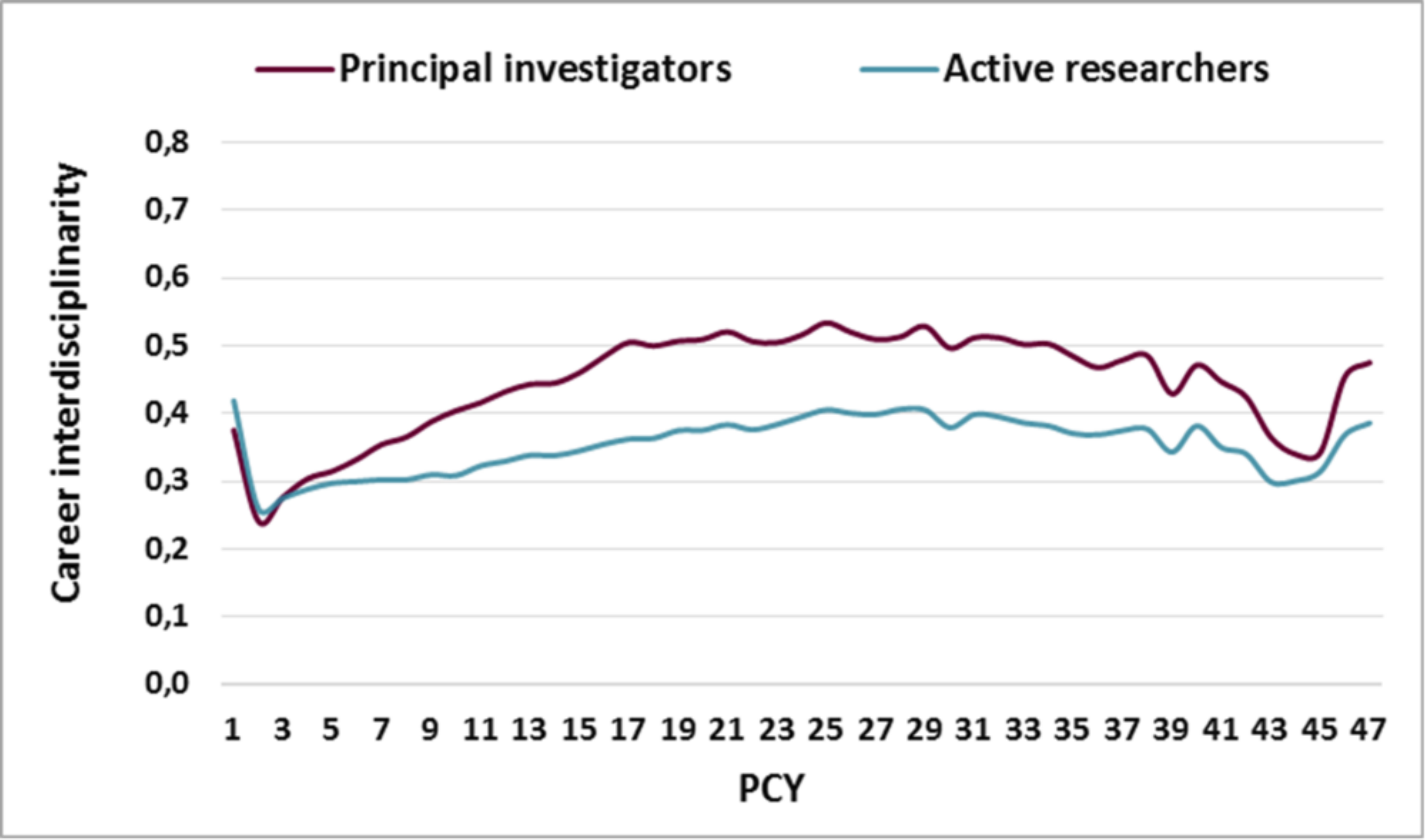}
\end{minipage}
\caption{Career internationality  (left) and interdisciplinarity (right) for PI's and for all active researchers. PI's have much more international and interdisciplinary career. We can observe similar dynamics but the intensity of internationality and interdisciplinarity is much bigger for PI's.}
\label{fig13:CareerInternatInterdisc}
\end{figure}

Career internationality and interdisciplinarity of PI's and of all active researchers are shown on Figure~\ref{fig13:CareerInternatInterdisc}. PI's have also much more international and interdisciplinary careers compared to all active researchers.

Note that the differences between PI's and all researchers are very small at the beginning (for $\PCY=1,2,3$) and at the end ($\PCY \ge 40$). The reason is that at the beginning they all have very similar conditions for work, while at the end the majority of active researchers are PI's.

\begin{figure}[htp!]
\centering
\includegraphics[width=.75\textwidth]{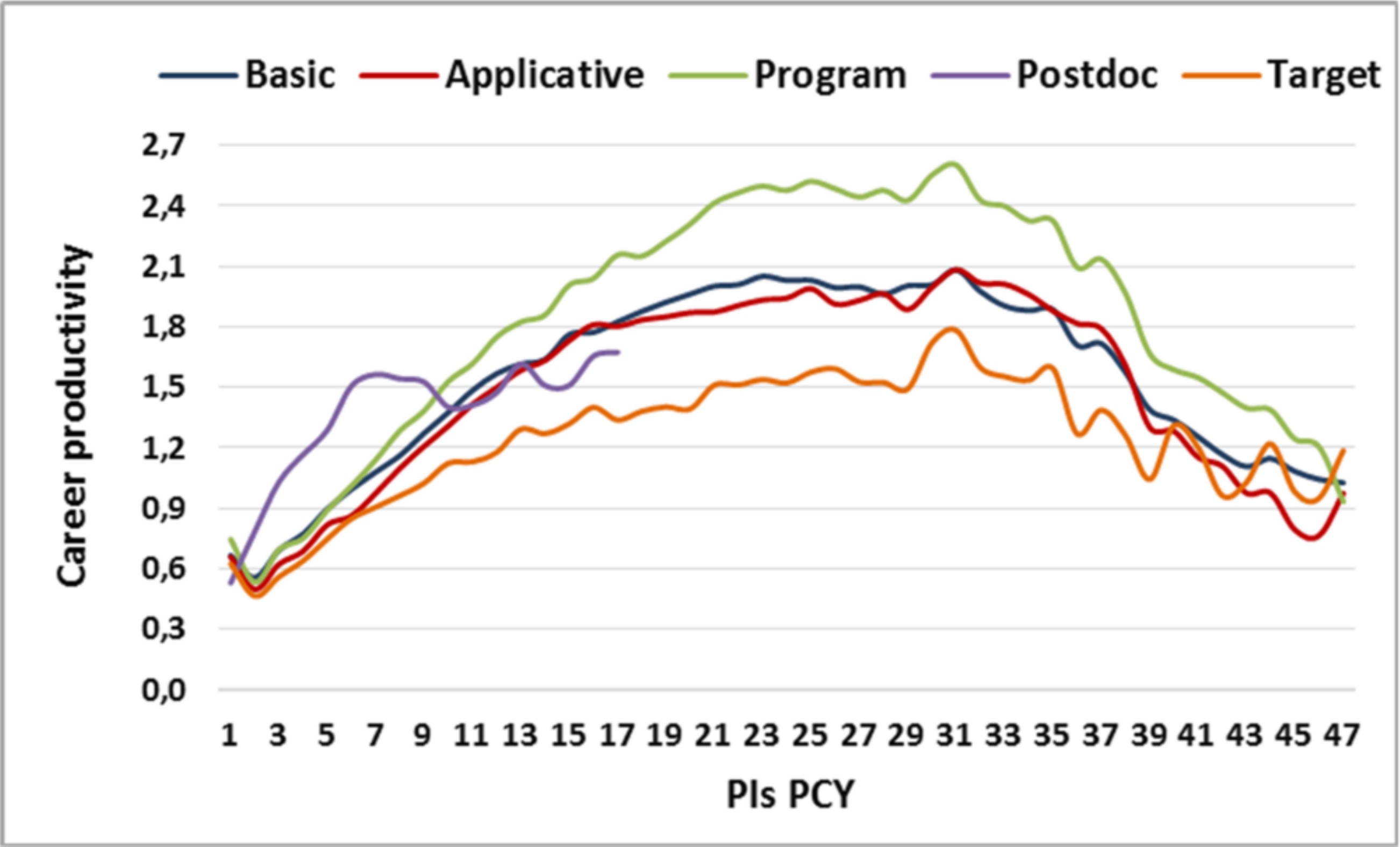}
\caption{The career productivities of different subgroups of $\PI$. We can see that at the beginning the postdoc PI's are very outstanding, but later in career path the best performing PI's become those that lead the prestigious research programmes.}
\label{fig14:PIsSubgroup}
\end{figure}

We end this section by demonstrating which subgroup of PI's is the most productive. Recall that there are several types of projects: postdoctoral, research programs, basic and applicative research projects and targeted research projects. Figure \ref{fig14:PIsSubgroup} depicts career productivities of PI's of different types of projects.

Note that the PI's that were granted in their (early part of) career also a postdoc project have at the beginning much better performance compared to the others. This is not surprising since in Slovenia a postdoc project means that the PI of such a project has complete control over the contents, dynamics and expenditure on such a project. However, in the 8th year after the first publication (this roughly coincides with the end of the postdoc project) their career productivity starts deviating towards the average career productivity of all PI's. This demonstrates that the postdoc projects are granted to young researchers that have outstanding scientific results in their early career, which are usually related to their PhD research. 

However, to become a leading researcher in their domain and to keep growing their research performance, the researchers must establish their own group, start new collaborations and provide additional funding for the group members, travelling and equipment. To achieve all these, additional skills and efforts are needed and winning postdoc grant does not guaranty them. Note that $70\%$ of researchers that won postdoc project before 2011 (hence have finished it before 2014) did not succeed to win another research project by 2017. 

\section{PI's are the backbone of collaboration network}

In this section, we investigate in detail what role do PI's play in the collaboration network of Slovenian researchers. Here, by an aggregate collaboration network in the period from the year $y_i$ to the year $y_j$ we mean a network $\N_{y_i}^{y_j}$ with the vertex set comprised of the researchers that published at least one scientific publication in the given period. A pair of vertices is connected if the corresponding researchers co-authored a scientific publication in the given period.

We consider the development of the network in yearly time frame. We construct aggregate collaboration networks for the periods starting in 1970 and ending in the years from 1994 to 2016. We hence analyze the properties of the networks $\N_{1970}^{1994}$, $\N_{1970}^{1995}$,\dots, $\N_{1970}^{2016}$. In particular, in Table~\ref{tbl:net}, we list several basic invariants: the sizes $n_A$ of the networks (i.e., the number of vertices), the number of different PI's in the network $n_\PI$, the relative size of the giant component $\gamma^{\mathrm{GC}}$ (i.e., the quotient between the number of vertices in the giant component and $n_A$), and the number of PI's in the giant component $n_{\PI}^{\mathrm{GC}}$. We also include several more involved invariants, which we describe below.

\begin{table}[htbp!]	
\begin{tabular}	{|c|c|c|c|c|c|c|c|c|c|c|}
\hline
& 	$n_A$ 	& $n_\PI$ 	& $\gamma^{\mathrm{GC}}$	& $n_{\PI}^{\mathrm{GC}}$	& $n_\PI^*$	& $\alpha^*$& $\nu_A$		& $\nu_{\mathrm{\PI}}$	& $\zeta^\times$			& $\gamma^\times$ 	\\ \hline
$\N_{1970}^{1994}$	& 	7084	& $22$ 	& 0.75	& $19$	& $19$	& $0$	& $/$	& $/$	& $25$ ($19$) 	& $0.99$	\\ \hline
$\N_{1970}^{1995}$	& 	7709	& $136$	& 0.77	& $109$	& $89$	& $0.07$	& $0.30$	& $0.11$	& $157$ ($143$) 	& $0.95$	\\ \hline
$\N_{1970}^{1996}$	& 	8338	& $557$	& 0.78	& $484$	& $374$	& $0.32$	& $0.34$	& $0.13$	& $822$ ($753$)	& $0.77$	\\ \hline
$\N_{1970}^{1997}$	& 	8939	& $876$	& 0.78	& $770$	& $280$	& $0.76$	& $0.47$	& $0.24$	& $1354$ ($1235$)	& $0.66$	\\ \hline
$\N_{1970}^{1998}$	& 	9620	& $1072$	& 0.79	& $942$	& $165$	& $0.85$	& $0.54$	& $0.33$	& $1654$ ($1516$)	& $0.62$	\\ \hline
$\N_{1970}^{1999}$	& 	10202	& $1267$	& 0.80	& $1115$	& $161$	& $0.87$	& $0.56$	& $0.37$	& $1908$ ($1742$)	& $0.59$	\\ \hline
$\N_{1970}^{2000}$	& 	10935	& $1367$	& 0.80	& $1219$	& $87$	& $0.93$	& $0.58$	& $0.39$	& $2091$ ($1907$)	& $0.58$	\\ \hline
$\N_{1970}^{2001}$	& 	11585	& $1593$	& 0.81	& $1433$	& $198$	& $0.87$	& $0.59$	& $0.42$	& $2324$ ($2120$)	& $0.56$	\\ \hline
$\N_{1970}^{2002}$	& 	12206	& $1691$	& 0.82	& $1536$	& $86$	& $0.81$	& $0.60$	& $0.44$	& $2512$ ($2307$)	& $0.56$	\\ \hline
$\N_{1970}^{2003}$	& 	12854	& $1753$	& 0.83	& $1606$	& $53$	& $0.83$	& $0.61$	& $0.44$	& $2688$ ($2469$)	& $0.56$	\\ \hline
$\N_{1970}^{2004}$	& 	13519	& $1900$	& 0.84	& $1751$	& $131$	& $0.96$	& $0.60$	& $0.44$	& $2885$ ($2668$)	& $0.55$	\\ \hline
$\N_{1970}^{2005}$	& 	14158	& $1935$	& 0.85	& $1809$	& $32$	& $0.97$	& $0.61$	& $0.44$	& $3061$ ($2838$)	& $0.56$	\\ \hline
$\N_{1970}^{2006}$	& 	14829	& $2022$	& 0.85	& $1903$	& $84$	& $0.89$	& $0.61$	& $0.44$	& $3213$ ($2977$)	& $0.56$	\\ \hline
$\N_{1970}^{2007}$	& 	15527	& $2149$	& 0.86	& $2029$	& $115$	& $0.93$	& $0.61$	& $0.46$	& $3328$ ($3097$)	& $0.57$	\\ \hline
$\N_{1970}^{2008}$	& 	16188	& $2257$	& 0.87	& $2144$	& $99$	& $0.98$	& $0.61$	& $0.47$	& $3461$ ($3253$)	& $0.57$	\\ \hline
$\N_{1970}^{2009}$	& 	16819	& $2363$	& 0.87	& $2254$	& $100$	& $0.94$	& $0.61$	& $0.48$	& $3658$ ($3447$)	& $0.57$	\\ \hline
$\N_{1970}^{2010}$	& 	17344	& $2445$	& 0.88	& $2342$	& $75$	& $0.96$	& $0.61$	& $0.49$	& $3771$ ($3570$)	& $0.58$	\\ \hline
$\N_{1970}^{2011}$	& 	17952	& $2524$	& 0.89	& $2426$	& $76$	& $0.97$	& $0.61$	& $0.49$	& $3916$ ($3722$)	& $0.58$	\\ \hline
$\N_{1970}^{2012}$	& 	18419	& $2528$	& 0.90	& $2435$	& $4$	& $0.75$	& $0.61$	& $0.48$	& $3973$ ($3785$)	& $0.59$	\\ \hline
$\N_{1970}^{2013}$	& 	18882	& $2573$	& 0.90	& $2486$	& $44$	& $0.91$	& $0.61$	& $0.47$	& $3986$ ($3809$)	& $0.60$	\\ \hline
$\N_{1970}^{2014}$	& 	19197	& $2619$	& 0.91	& $2534$	& $44$	& $0.93$	& $0.60$	& $0.45$	& $3996$ ($3831$)	& $0.61$	\\ \hline
$\N_{1970}^{2015}$	& 	19452	& $2650$	& 0.91	& $2566$	& $30$	& $0.93$	& $0.60$	& $0.45$	& $3969$ ($3804$)	& $0.62$	\\ \hline
$\N_{1970}^{2016}$	& 	19598	& $2725$ 	& 0.91	& $2637$	& $70$	& $0.99$	& $0.60$	& $0.46$	& $3932$ ($3752$)	& $0.62$	\\ \hline
\end{tabular}
\caption{The list of computed invariants for the aggregated collaboration networks: the number of vertices ($n_A$), the number of PI's ($n_\PI$), the relative order of the giant component ($\gamma^{\mathrm{GC}}$), the number of PI's in the giant component ($n_{\PI}^{\mathrm{GC}}$), the number of first time PI's in the last year of the given period ($n_\PI^*$), the proportion of first time PI's which are connected to some previous PI ($\alpha^*$), the average relative number of PI's in researcher's ego-network ($\nu_A$), the average relative number of PI's in PI's ego-network ($\nu_{\mathrm{\PI}}$), the number of connected components when PI's are removed from the giant component and the number of components containing only one vertex ($\zeta^\times$), and the relative size of the giant component after the PI's are removed ($\gamma^\times$).}
\label{tbl:net}
\end{table}

As customary in the network analysis, we will be mainly interested the dynamics in the \textit{giant component}, i.e., the largest connected component of a network. In the year 1994, the giant component of $\N_{1970}^{1994}$ is comprised of $75\%$ of all vertices, and is constantly growing through the years, containing more than $91\%$ of vertices in the year 2016. In fact, ignoring the researchers that have no co-author among active researchers in the observed period, the giant component of $\N_{1970}^{2016}$ consists of more than $99\%$ of researchers. Hence, it is not a surprise that almost all PI's also belong to the giant component---$86\%$ of all PI's in the year 1994 and $97\%$ in the year 2016. However, there are $88$ PI's in 2016, which are not a part of the giant component.

In the column $n_\PI^*$ of Table~\ref{tbl:net}, we list the number of PI's in the giant component that have been granted a project for the first time in the last year of the period covered by a particular network. One may observe that their number is very vibrant, but we have, however, detected an extremly interesting and stable property regarding them. It is highly likely that a PI being granted a project for the first time is connected in the network to a PI from previous years. In fact, the proportion $\alpha^*$ of such PI's is strongly above $90\%$ since 2004. In our opinion, this demonstrates the (well-known) fact that a researcher can easier win a grant if somebody in her neighbourhood motivates her to start writing a proposal and also mentors her how to write a good proposal.

An ego-network for a vertex $r$ is a subnetwork induced on $r$ and all its neighboring vertices. In our setting, the ego-network of a researcher is comprised of all her co-authors within the observed period. For every researcher $r$, we have computed the relative number of PI's in her ego-network as the quotient of the number of PI's and the number of all collaborators in her ego-network. We computed the average of these quotients over all active researchers in the giant components ($\nu_A$) and of all PI's in the giant component ($\nu_{\mathrm{\PI}}$).
 
The results show that PI's are well distributed across the networks. Although their number is below $20\%$ of all vertices, every researcher has about $60\%$ of PI's in her ego-network already since 2000. The proportion is, at first sight surprisingly, smaller for PI's. Every PI has on average about $45\%$ of PI's in her ego-network. This indicates high collaboration between PI's, but the surprising fact is that the average is higher in general (i.e., $\nu_A>\nu_{\mathrm{\PI}}$.). We drilled into the underlying data and figured out that the difference is mostly brought in by the researchers having small neighbourhoods (less than three co-workers in their ego-network). Many such researchers are connected only to PI's (being their advisors), and hence have the proportion equal to $1$. Ignoring such researchers in our computations, the average proportion never exceeds $50\%$, but it is still constantly higher than $\nu_{\mathrm{\PI}}$.

The above discovery suggests another aspect to consider when analyzing positions of PI's in bibliographic networks. Evidently, they are connecting the network. Recall that the size of the giant component is more than $91\%$ in the last years of the observed period. So, what happens if one removes PI's from the network? It turns out that the networks break into many smaller connected components, mostly of size $1$ (which are most likely PIs' students). We list the number of those components in the column $\zeta^\times$ with the number of single-vertex components in the brackets. After removing PI's, there does however retain a giant component comprised of majority of all vertices, but its size $\gamma^\times$ regarding the whole network is just about $60\%$, which is rather expected. This implies a conclusion that the PI's indeed form a backbone of the networks.

High interconnectivity between PI's can also be observed by constructing aggregate collaboration networks induced only on the vertices which have been PI's in the observed period. More precisely, we limited ourself to subnetworks of $\N_{1970}^{1994}-\N_{1970}^{2016}$ spanned by giant components. In each of them we considered the subnetworks spanned only by PI's. Again, the giant component of the ``PI-network'' of 2002 is already comprised of over $90\%$ of vertices and this percentage is later constantly growing. The number of connected components (mostly isolated researchers) is the highest in 2001 and then, interestingly, constantly decreasing. This means that either isolated PI's are connecting with other PI's or their co-authors became PI's during the years.

\begin{figure}[htp!]
\centering
\includegraphics[width=.75\textwidth]{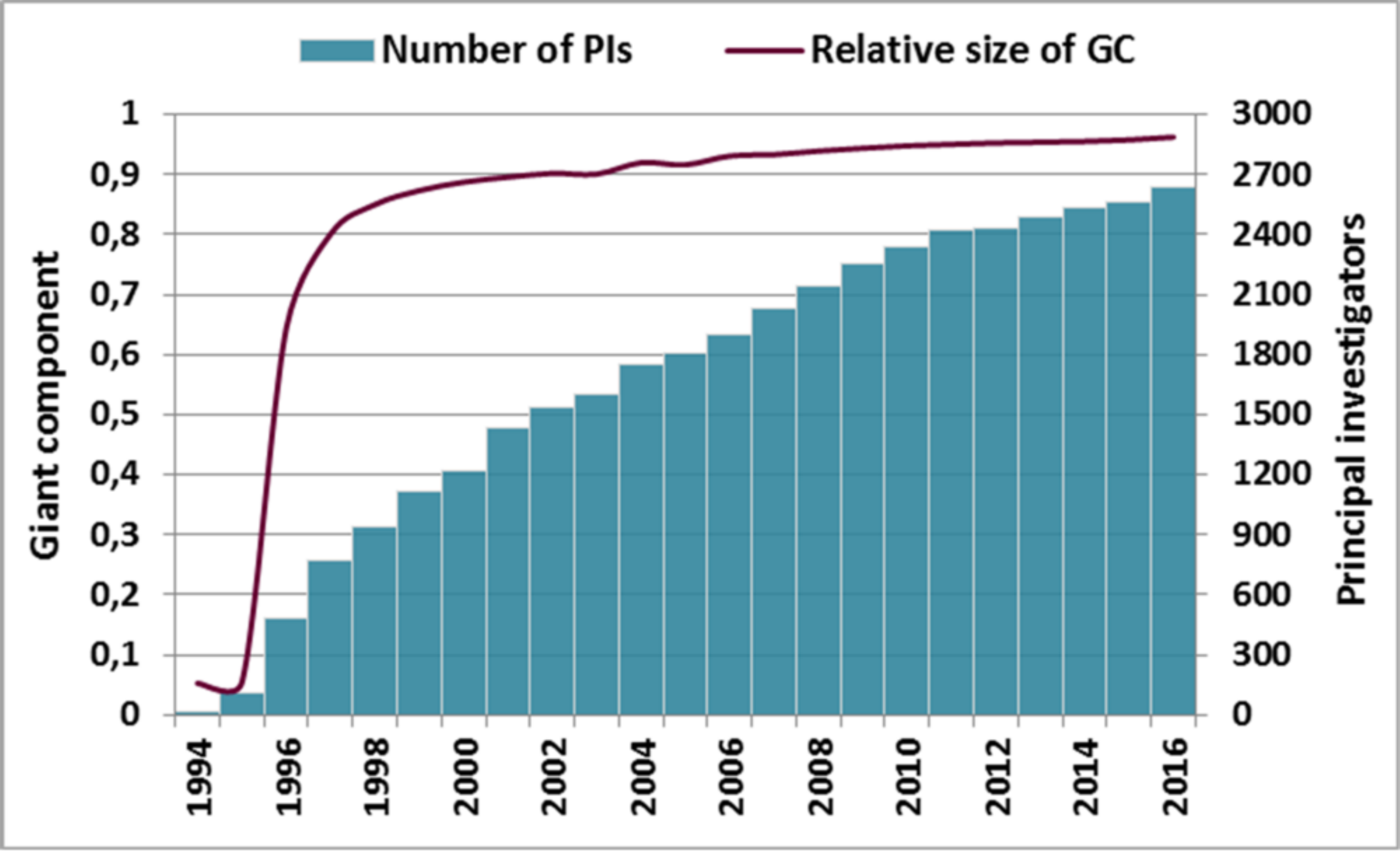}
\caption{The aggregate collaboration network of principal investigators}
\label{fig16:PIsNet}
\end{figure}

\section{Conclusions}

In this paper we compared scientific performance of (i) Slovenian researchers, that have succeeded to win at least one publicly funded research grant in Slovenia since 1994 (we call them PI's) and of (ii) all active researchers---those that were registered in SiCRIS in June 2017 and have published at least one scientific publication in the period 1970--2016. 

The comparison was based on four main aspects of scientific performance including productivity, collaboration, internationality and interdisciplinarity. Our analyses are based on high quality data about researchers, publications and  national research grants. 

In the first part of the paper we found out that the PI's are outperforming the average researchers in all $4$ categories: they publish (much) more scientific publications, have larger sets of collaborators and more international co-authors on their scientific publications and are more interdisciplinary. Excellent research capabilities are crucial for PI's to be successful in leading scientific projects~\citep{Cunningham2016a}. This is also in line with recent findings proposed by \citet{Cunningham2018} who argue that becoming a PI transforms individual scientific careers and their trajectories.

Additionally, we compared the career paths of both groups of researchers. In order to be able to compare careers of researchers from different time periods we introduced career productivity, career collaboration, career internationality, and career interdisciplinarity. 

Our analysis showed that PI's have longer and more fruitful academic careers. More precisely, an average PI is active $26.5$ years (the average period between the first and the last publication), while all active researchers are on average active $14.1$ years. During their careers PI's have higher values for all four career indicators. When we considered PI's of different types of projects we figured out that the most productive careers pertain to the PI's of research programmes, which are the most prestigious national grants. Similar findings were proposed by \citet{Feeney2014}. They demonstrated that PI's who receive grants, produce significantly more papers and supervise more students than those who do not receive grants.

PI's who were granted postdoc project at the beginning of their careers demonstrate  outstanding scientific performance at that time. However, after finishing their project, their career productivity starts deviating towards the average career productivity. This is is line with ``utility maximizing theory'' according to which scientists reduce their research efforts over time~\citep{Kwiek2015}. Opposite evidence were presented by \citet{Yang2015} who claimed that completing a postdoctoral position positively correlates with working in academia and also secure tenure-track appointment. Additionally, $70\%$ of the researchers who won postdoc project before 2011 (hence have finished it before 2014) did not succeed to win another research project by 2017. In our opinion these finding suggest that better nurture of these young people is needed if we want them to grow from outstanding young researchers to leading researcher in their domain.

In the last part of the paper we analyzed the collaboration networks spanned by the Slovenian researchers that had been active in the periods 1970--1994, 1970--1995, \ldots, 1970--2016. We have detected several interesting patterns. Firstly, these networks are well connected (the size of the giant component grows above $90\%$ in the last years of observation) and the PI's are well distributed across them. Additionally, by removing the PI's from these networks, the networks break into many smaller connected components, mostly of size $1$. However, there does retain a giant component comprised of majority of all vertices, but its size regarding the whole network is just about $60\%$, which is rather expected. The PI's are indeed a backbone of these networks. Additionally, a researcher that becomes a PI for the first time has in most cases (in more than $90\%$) a PI in her collaboration network. We also found out that even collaboration networks spanned exclusively by PI's very soon form giant components which consist of more than $90\%$ of the network members. This additionally highlights the importance of PI's, but also calls for caution to avoid potential negative impacts of such situation.

All these results confirm that in the scientific community of Slovenia the PI's have a central role. By demonstrating outstanding scientific performance and having central position in the scientific network, they are certainly a very important driver of science in Slovenia. This needs to be more widely acknowledged. There are however some questions which still remain unanswered. For instance, it would be interesting to see, how PI's perform as advisors in comparison to other researchers. Are their students more successful? Another question is if the PI's are the ones who are establishing new directions in science in sense of being initiators of new research fields.

\bibliographystyle{spbasic}
\bibliography{biblio}
\end{document}